\documentclass{osa-article}

\journal{osajournal}
\articletype{Research Article}
\usepackage{color,soul}

\begin{document}

\title{Modelling the Partially Coherent Behaviour of Few-Mode Far-Infrared Grating Spectrometers}

\author{B.N.R. Lap,\authormark{1,2,*} S. Withington ,\authormark{3}, W. Jellema,\authormark{2,1} and D.A. Naylor \authormark{4}}

\address{\authormark{1}Kapteyn Astronomical Institute, University of Groningen, 9700 AV, Groningen, The Netherlands\\
\authormark{2}SRON Netherlands Institute for Space Research, 9700 AV, Groningen, The Netherlands
\authormark{3}Cavendish laboratory, JJ Thomson Avenue, Cambridge CB3 OHE, UK\\
\authormark{4}Institute for Space Imaging Science, Department of Physics and Astronomy, University of Lethbridge, 4401 University Drive, Lethbridge, Alberta T1K 3M4, Canada}

\email{\authormark{*} b.lap@sron.nl} %% email address is required
%% To be edited by editor
% \dates{Compiled \today}

%\ociscodes{(140.3490) Lasers, distributed feedback; (060.2420) Fibers, polarization-maintaining;(060.3735) Fiber Bragg gratings.}

%% To be edited by editor
% \doi{\url{http://dx.doi.org/10.1364/XX.XX.XXXXXX}}

\begin{abstract}
Modelling ultra-low-noise far-infrared grating spectrometers has become crucial for the next generation of far-infrared space observatories. Conventional techniques are awkward to apply because of the partially coherent form of the incident spectral field, and the few-mode response of the optics and detectors. We present a modal technique for modelling the behaviour of spectrometers, which allows for the propagation and detection of partially coherent fields, and the inclusion of straylight radiated by warm internal surfaces. We illustrate the technique by modelling the behaviour of the Long Wavelength Band of the proposed SAFARI instrument on the well-studied SPICA mission.
\end{abstract}

\section{Introduction}
The next generation of far-infrared (FIR), 30 – 300\,$\mu$m, space-based astronomical telescopes will use cooled primary optics (< 4\,K) and ultra-low-noise superconducting detectors with Noise Equivalent Powers (NEP) of $< 10^{-19} \text{ WHz}^{-1/2}$ \cite{Jackson11} to achieve unprecedented levels of observing sensitivity. A typical payload will comprise a complex package of spectrometers, polarimeters and imaging photometers, and the behaviour of these instruments must be understood, both individually and collectively, with a high degree of confidence to ensure that the science goals of the mission are met.

One of the key science goals is to understand the details of star and planet formation in our own and extrasolar planetary systems \cite{Roelfsema:2018,Leisawitz:18,Duncan:19}. In order to do this, astronomers need to extract statistically meaningful spectroscopic information from young stars and their protoplanetary disks to determine the composition, distribution and quantity of the gas, dust and more complex molecules, such as water \cite{kamp:21}. Ideally, this information would also include a spatial map, but the angular resolving power of future FIR facilities will be limited, due to the prohibitive cost of cold, monolithic, primary mirrors. As a result, the majority of these observatories will try to disentangle spatial information using a combination of broad-band, low, medium and high spectroscopic measurements of what are, effectively, point sources.

Broad-band low-resolution spectroscopy ($R\sim100$) is best achieved by using a grating spectrometer (GS), while broadband medium-resolution spectroscopy ($R\sim1000$) of point sources is best achieved using a Fourier Transform Spectrometer (FTS). These instruments typically feed a spatial-spectral array of ultra-sensitive photometric detectors. To achieve broad-band high-resolution spectroscopy ($R\sim10^4 - 10^5$) in ultra-low-noise systems, it is necessary to keep background loading and the photon noise to a minimum, i.e. by limiting the spectral band. In FIR space borne astronomical spectrometers, this can be accomplished by post-dispersing the light from an FTS using a diffraction grating, i.e. a Post Dispersed Fourier Transform Spectrometer (PDFTS) \cite{Wiedemann:89,Hajian:07}, before it is fed onto an ultra-low-noise detector array.

At long FIR wavelengths, it is generally beneficial for the individual detectors in the focal plane of a spectrometer to be few-mode (5-20), enabling an increase in the overall system throughput, but at the cost of increased coupling to straylight and thermal background radiation, as seen in the in-flight behaviour of the Herschel-SPIRE instrument \cite{Makiwa:13,Swinyard:14,Valtchanov:17}. The next generation of telescopes, as typified by the SPICA \cite{Roelfsema:2018},  OST \cite{Leisawitz:18}, and GEP \cite{Glen:21} missions, will be at least two orders of magnitude more sensitive than Herschel-SPIRE. Therefore, low-level artefacts similar to the ones seen in Herschel-SPIRE will be even more prominent, but also additional issues will arise, which were not encountered with SPIRE due to its lower observing sensitivity. To ensure the success of future FIR spectroscopic missions both of these unknowns must be addressed and controlled prior to launch with a high degree of precision. 

From a design perspective, it will be necessary to tailor the spatial-spectral response of few-mode instruments, or equivalently to design the optics and detectors using the language of partially coherent optical fields, rather than simply accepting the single-mode (fully coherent) or multi-mode (fully incoherent) extremes. This shift in mindset is particularly pronounced when considering matters such as how best to scale the size of optical components (e.g entrance and exit slits, baffles, grating geometry etc.); how best to pack and sample the focal plane with typically hundreds of ultra-sensitive detectors; how best to engineer the few-mode response of the individual detectors; how best to achieve experimental characterisation and verification; and how best to calibrate the partially coherent behaviour.  

In order to address these issues, an accurate optical model of FIR spectrometers is required. We could opt for fully coherent (e.g. full electromagnetic simulations \cite{Davidson:04}), fully incoherent (e.g. geometrical optics \cite{BornWolf:99}), or alternative modelling techniques such as Fourier Optics \cite{Goodman:05} and Gaussian-Laguerre mode decomposition \cite{Goldsmith:98,Sullivan:09}. Although these classical approaches provide valuable insights, they do not account for few-mode partial coherence and associated behaviour. In fact, we are not aware of a comprehensive simulation method that can handle matters such as partially coherent analysis and design, background power loading calculations, and straylight analysis of few-mode FTS and GS, let alone the much more complicated PDFTS. 

We have developed a numerical procedure for simulating the optical behaviour of ultra-sensitive few-mode FIR instruments. The method is based on a continuous functional theory of grating spectrometers \cite{Withington:21}, and comprises four steps: i) the second-order spatial correlation function of the incident electromagnetic field is established; ii) the correlation function is propagated through to the output plane using the optical modes of the system, including dispersive components such as gratings; iii) the state of coherence of internally generated thermal radiation is calculated, and combined with the signal; iv) the total partially coherent field is coupled to the state of coherence to which the detectors are sensitive, yielding, for example, the power recorded by each pixel as a function of wavelength. This approach is numerically powerful, allowing a wealth of behaviour to be studied in a way that is conceptually meaningful.

The purpose of this paper is i) to present a numerical implementation of the continuous function theory used for describing few-mode optics; ii) to explore the behaviour of the method using a one-dimensional (1-D) GS as an illustrative example; iii) to ensure that the method gives intuitive results in those cases where behaviour can be predicted; (iv) to gain a conceptual appreciation of the operation of few-mode GS, including some trade-offs that are relevant when optimising a design. Although we have carried out simulations of both FTS and GS, we will use the SPICA/SAFARI Long Wavelength Band grating spectrometer as a case-study, because it is representative for the next generation of few-mode grating spectrometers. In later papers, we shall present the full functional model, and describe the application of the numerical techniques to more complicated spectrometer designs.

\section{Theory}
\label{sec:theory}

This section is comprised of five parts. In the first part, a generic GS scheme, and a representative 1-D GS optical model are introduced. In the second part, the incident electric field over the slit is introduced, and we examine its state of coherence. In the third part, we move on to the grating module optics, where we discuss the details of the numerical modal framework; how it provides the optical modes; how a spectrum is measured; and how straylight can be included into the framework. In the fourth part, we discuss the detector array, and in the last part, the fifth part, we will integrate the detector array with the grating optics module and provide a polychromatic description of a few-mode GS. 
 
\subsection{Grating spectrometer optical model}
\label{sec:GS_optical_model}

In a space-based GS, a system of fore optics focuses the electromagnetic field incident on the telescope onto an entrance of a vertical slit. The resulting diffracted field is passed through collimating optics, after which it is dispersed by a grating before arriving at the camera optics. At this stage, the dispersed field is focused onto the focal (or output) plane, where a two-dimensional array of photometric detectors measures the total power. This detector array can have a spatial direction, in addition to the spectral direction, to allow angular cuts through extended astronomical sources.

In this paper, we will use the SAFARI Long Wavelength Band design as a case study, but before we discuss this GS model, we need to explain the underlying assumptions of the simulations. These assumptions are not intrinsic to the developed framework, and numerous extensions are possible, but they set the boundaries within which the simulations were carried out.
First, we consider linear physical optics, and will explicitly ignore polarization effects, i.e. we will consider linearly polarised electric fields only. 
Second, we will consider optical elements that are optically thin, such that refraction and reflection within the optical components can be ignored. Although the optically thin limit is an approximation, it is commonly used and valid at FIR wavelengths \cite{Goldsmith:98}.
Third, and finally, we ignore the fore optics, to focus on the behaviour of the GS. 

\begin{figure}[t!]
\centering
\includegraphics[width = 8.4 cm]{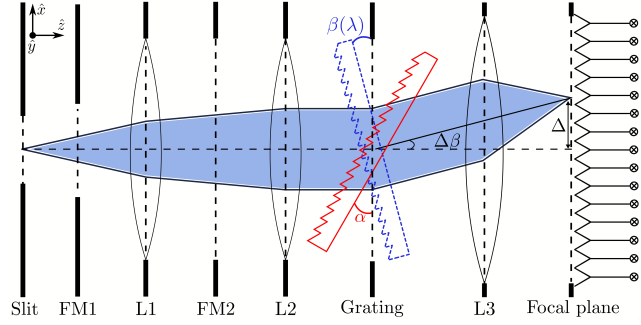}
\caption{ Cross section of the in-line equivalent SAFARI Long Wavelength Band model in the $xz$ plane, where $\hat{z}$ is the axis of propagation. The diffracted, partially coherent incident field enters the GS at the (entrance) slit with some focal ratio, $F$, and gets reflected by the first fold mirror ($\text{FM1}$). The in-line equivalent of a fold mirror is an aperture. Then, the field is converged by the first mirror ($\text{L1}$), which is represented by a thin lens. The beam then reaches a second fold mirror ($\text{FM2}$) and gets collimated by the second mirror ($\text{L2}$). It arrives at the grating under an angle, $\alpha$, gets dispersed and exits under an angle, $\beta(\lambda)$. Then, the dispersed radiation is focused onto the focal plane by the camera lens ($\text{L3}$), where a linear bolometric detector array, defining the exit slit of the GS, is placed to measure the power of the incident partially coherent field. See the text for the definitions of $\Delta \beta$ and $\Delta$.}
\label{fig:Lap_fig1}
\end{figure}

In this paper, we focus on modelling the behaviour of a GS in the spectral direction, We separate its spectral behaviour from its behaviour in the spatial dimension as follows.
First, we define an in-line equivalent model using the three-dimensional optical model of the SAFARI Long Wavelength Band. In this in-line model, the spectral and spatial dimension of the GS are defined along the $\hat{x}$ and $\hat{y}$ axis, respectively, and the optical axis is defined along $\hat{z}$, which coincides with the propagation axis of the incident electric field over the slit. Furthermore, the 3-D optical surfaces (such as fold mirrors, mirrors with optical power, and the diffraction grating) are replaced by their in-line optically thin equivalents (i.e. apertures, lenses, and a transmission grating), which are defined in the $xz$ plane. 
Second, we assume that the in-line optical elements of the GS extend to infinity in the spatial direction, i.e. along $\hat{y}$. Under this assumption, we can isolate the behaviour of the GS in the spectral dimension by taking a cross cut of the in-line model in the $xz$-plane. The resulting 1-D GS model is shown in Fig. \ref{fig:Lap_fig1}.

Three observations observations about the 1-D GS model are worth noting. 
First, two modifications are made to the basic GS layout. Namely, two fold mirrors ($\text{FM1}$ and $\text{FM2}$) are added to minimise the volume of the three-dimensional instrument, and the lenses $\text{L1}$ and $\text{L2}$ are added to control anamorphic magnification.  
Second, the front and back surface of the grating are inclined planes, as indicated in Fig. \ref{fig:Lap_fig1} by the solid red and dashed blue lines, respectively. The inclination of these planes are controlled by the angle of incidence, $\alpha$, which in this 1-D model is constant, and angle of reflectance, $\beta(\lambda)$. It is convenient to define a reference wavelength, $\lambda_0$, and a reference angle of reflectance, $\beta_0 = \beta(\lambda_0)$, such that $\Delta \beta = \beta(\lambda ) - \beta_0$. Under this definition, when the operation wavelength $\lambda = \lambda_0$, the transverse direction of the propagating field is unchanged and the spatial displacement of the beam over the focal plane with respect to the optical axis (the horizontal dashed line in Fig. \ref{fig:Lap_fig1}), $\Delta = 0$. 
Third, and finally, diffraction gratings are generally used in reflection, but for simplicity an (ideal) transmission grating is shown. In reality multiple reflection and absorption components are to be expected, but here these higher order contributions will be ignored and we primarily focus on first order effects. Moreover, the transmission grating is used under an high angle of incidence to achieve a grating resolution, $R \sim300$. In practice, for high $\alpha$, a diffraction grating becomes a polarization filter, because the polarization perpendicular to the direction of the grating grooves is diffracted with higher efficiency. This polarization sensitivity can be exploited using a Martin-Puplett FTS \cite{Martin:70}, resulting in an increase in observing sensitivity, because the photon background is suppressed. However, as stated previously we ignre polarization effects. 

The optical modelling is simplified when the polychromatic behaviour of the 1-D GS can be considered to be a collection of mutually incoherent monochromatic realizations. In other words, it would be beneficial to first model the GS at discrete frequencies, and then later on to combine the monochromatic simulation results to provide a polychromatic description of the system as a whole. This desideratum is met when: i) complex statistically stationary electric fields are considered, and ii) the random fluctuations in the fields are narrow in bandwidth $\Delta \nu$ compared its the mean frequency $\bar{\nu}$, i.e. \cite{Wolf:07}
\begin{equation}
    \frac{\Delta \nu}{\bar{\nu}} << 1.
    \label{eq:theory_requirement}
\end{equation}
In astronomical telescopes, these conditions are met. 

In what follows, we will describe the details of each numerical step for monochromatic light, and then combine the results to provide a description of the complete grating spectrometer.

\subsection{Field distribution over the slit}
\label{sec:field_distribution_over_slit}

A central assumption is that the slit on the input surface of the GS is sufficiently small such that the spatial state of coherence of the electric field over the slit does not vary appreciably with frequency. If this were not the case, the partially coherent field could not be regarded as having an overall well-defined spectrum, because the spectrum may vary with position.

Under this assumption, an incident electric field over the slit, at a single discrete frequency, $\nu$, can be sampled and written as a column vector
\begin{equation}
    \boldsymbol{\rm{e}} = \big[ e_1, e_2,\dots, e_N \big]^{T} ,
    \label{eq:theory_input0}
\end{equation}
where the sample positions are given by  
\begin{equation}
	\boldsymbol{\rm{x}}^{(slit)} = \big[ x_1^{(slit)},x_2^{(slit)}, \dots , x_N^{(slit)} \big]^{T}.
	\label{eq:theory_input1}
\end{equation}
Here, $^{(in)}$ labels the slit, $x_{n}^{(in)}$ is the $n$-th element of $\boldsymbol{\rm{x}}^{(in)}$ with $n=1,2,...,N$, where $N$ is the total number of sample points, and $^{T}$ is the transpose. The spatial state of coherence of electric field $\boldsymbol{\rm{e}}$ at the sample points is given by the (spatial) correlation matrix
\begin{equation}
	\boldsymbol{\rm{E}} = \big< \boldsymbol{\rm{e}} \boldsymbol{\rm{e}}^{\dagger} \big> ,
	\label{eq:theory_input2}
\end{equation} 
where $^{\dagger}$ indicates the Hermitian transpose, and $\big< \, \, \big>$ indicates averaging over a representative ensemble \cite{Wolf:82,Withington:01,Withington:07}. Furthermore, the total power in the field is proportional to the trace of matrix $\boldsymbol{\rm{E}}$, and from now on lower case will be used for vectors and upper case for matrices.

In general, the electric field $\boldsymbol{\rm{e}}$ can be written as a weighted linear combination of individual fully coherent, but mutually fully incoherent, fields, which we shall refer to as modes. The eigenvectors of the Hermitian matrix $\boldsymbol{\rm{E}}$ give the sampled spatial forms of these modes, and the eigenvalues give the individual propagating powers. Electric field $\boldsymbol{\rm{e}}$ is partially coherent when its state of coherence lies between the two extremes of (spatial) coherence, and is said to be few-mode. In this case, matrix $\boldsymbol{\rm{E}}$ contains a finite set of eigenvectors.

To study the combined partially coherent behaviour of a few-mode FIR GS, and the impact it has on its performance, we need to model how a partially coherent field $\boldsymbol{\rm{E}}$ is propagated through the few-mode optical system, including the grating, onto the output plane, where it is coupled to the modes to which the detectors are sensitive.

\subsection{Grating module optics}
\label{sec:theory_GM_optics}

Modal optics, which relies on continuous functional theory, is an effective technique for describing few-mode optics at FIR wavelengths, due to two reasons. First, it can account for diffraction, whilst not having to evaluate multiple diffraction integrals to propagate a field through a complex optical system. Second, it avoids the need for full electromagnetic simulations, which are computationally intensive \cite{Withington:01}.

In modal optics, the notion of modes is used to map an incident electric field $\boldsymbol{\rm{E}}$, which can be in any state of coherence, over the input surface to the output surface of the optical system \cite{Withington:01,Withington:04,Withington:07}. This set of modes is characteristic for the optics, and is equivalent to a field propagator used for propagating incident fields through the optics. The Huygens-Fresnel Modal Framework (HFMF) approach we adopt in this paper is a numerical version of the functional theory \cite{Withington:01}, which uses the Huygens-Fresnel diffraction integral for obtaining this field propagator. The HFMF adopts the matrix notation of \cite{Ozaktas:02,Withington:04}, and as a result, the continuous field propagator is represented by the system transformation matrix, $\boldsymbol{\rm{H}}$.

Consider an optical system consisting of $S$ optical surfaces, with $s=1,2,...,S$ labeling the optical surfaces. Then, moving from the input surface, through the optics, to the output surface of the optical system, the matrix $\boldsymbol{\rm{H}}$ is given by:
\begin{equation}
    \boldsymbol{\rm{H}} =  \boldsymbol{\Theta}^{S} {\displaystyle \prod_{s=S-1}^{1}} \big(  \boldsymbol{\rm{T}}^{(s)} \boldsymbol{\Theta}^{(s)} \big) .
    \label{eq:theory_optics}
\end{equation}
Here, the indexing of the matrices is intentionally reversed to respect the ordering of the optical surfaces in accordance with the adopted matrix notation. In Eq. (\ref{eq:theory_optics}), the transmission matrices, $\{ \boldsymbol{\Theta}^{(s)} | s=1,2,...,S\}$, describe the phase transforming properties of the optical surfaces, and the propagation matrices, $\{ \boldsymbol{\rm{T}}^{(s)} | s=1,2,...,S-1 \}$, describe the mapping of the electric field over an optical surface to the next optical surface.

In words, Eq. (\ref{eq:theory_get_G1}) is interpreted as follows. For every $s$-th optical surface within the optical system, we determine i) the transmission matrix $\boldsymbol{\Theta}^{(s)}$ of that optical surface, and ii) the propagation matrix $\boldsymbol{\rm{T}}^{(s)}$ describing the mapping to the next optical surface with index $s+1$. This process is repeated until we reach the last optical surface with index $s=S$. At surface $S$ the optical system terminates, therefore, in Eq. (\ref{eq:theory_get_G1}), $\boldsymbol{\Theta}^{S}$ is the last matrix in the iterative sequence. 

The matrix $\boldsymbol{\rm{H}}$ describes the system propagation characteristics at a single discrete wavelength, and by applying an Singular Value Decomposition (SVD) to this matrix we can obtain the modes of the optical system. Before we do this for the 1-D GS model shown in Fig. \ref{fig:Lap_fig1}, we first turn to a detailed description of the propagation and the transmission matrices, and how they are obtained.

%%%%%%%%
\subsubsection{Propagation matrices}
\label{sec:propagation_matrices}

\begin{figure}[t!]
\centering
\includegraphics[width = 8.4 cm]{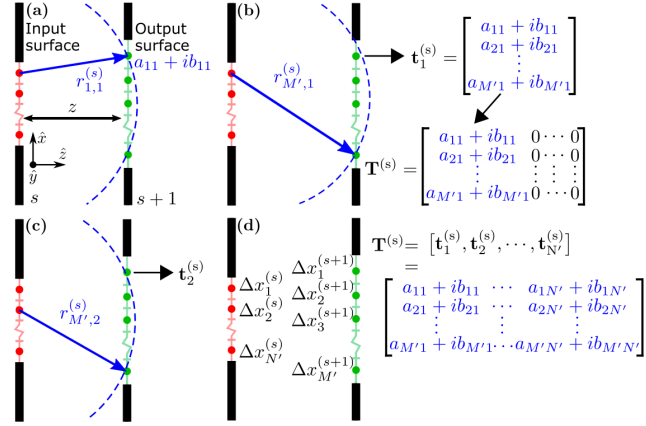}
\caption{Obtaining propagation matrix $\boldsymbol{\rm{T}}^{(s)}$ for a simple optical system, which comprises two optical surfaces labeled $s$ and $s+1$ of the same geometrical dimensions, and they are separated by a distance $z$. The $x,z$ coordinates of the input and output surface are sampled at $N'$ and $M'$ discrete positions in the $\hat{x}$ and $\hat{z}$ direction, which are $x_{n'}^{(s)}$ and $x_{m'}^{(s+1)}$, and $z_{n'}^{(s)}$ and $z_{m'}^{(s+1)}$, respectively, with $n'=1,2,\dots,N'$ and $m'= 1,2,\dots,M'$. Each discrete point has a sample step size, $\Delta x_{n'}^{(s)}$ and $\Delta x_{m'}^{(s+1)}$, as shown in (d). First, a spherical wave is emitted from the first discrete point over the entrance aperture ($n'=1$). The complex field value (indicated in blue) is determined for each point over the exit aperture, where in (a) $m'=1$ and in (b) $m'=M'$. The resulting complex field vector is stored column-wise in $\boldsymbol{\rm{T}}^{(s)}$. This process is repeated for each point over the entrance aperture (see (c)) until $\boldsymbol{\rm{T}}^{(s)}$ is fully populated (see (d)).}
\label{fig:Lap_fig2}
\end{figure}

To explain how the propagation matrices are obtained, we consider a simple optical system (see Fig. \ref{fig:Lap_fig2}(a)). This system is comprised of two optical surfaces: an input and an output surface, labeled $s$ and $s+1$, respectively. These surfaces have the same geometrical dimensions, act as apertures, and are separated by a distance, $z$. Moreover, the matrix $\boldsymbol{\rm{T}}^{(s)}$ is the propagation matrix of this system. 

We sample the input and output surface of this optical system using $N'$ and $M'$ discrete points in the $\hat{x}$ direction (see Fig. \ref{fig:Lap_fig2}). These sample points are stored in column vectors $\boldsymbol{\rm{x}}^{(s)}$ and $\boldsymbol{\rm{x}}^{(s+1)}$, with $n'=1,2,\dots,N'$ and $m' = 1,2,\dots,M'$, and at each discrete sample point a point source is placed. Each point source has a sample step size, i.e. $\Delta x_{n'}^{(s)}$ or $\Delta x_{m'}^{(s+1)}$ (see Fig. \ref{fig:Lap_fig2}(d)), which is $\leq \lambda/2$ with $\lambda$ being the operation wavelength, to ensure Nyquist sampling of the electric fields over the optical surfaces. Furthermore, each element in $\boldsymbol{\rm{x}}^{(s)}$ and $\boldsymbol{\rm{x}}^{(s+1)}$ has an associated $z$ position, which are stored in the column vectors $\boldsymbol{\rm{z}}^{(s)}$ and $\boldsymbol{\rm{z}}^{(s+1)}$, respectively, such that form an ordered pair with their sampled $x$ positions: $\big[ \boldsymbol{\rm{x}}^{(s)} ,\boldsymbol{\rm{z}}^{(s)} \big]$ and $\big[ \boldsymbol{\rm{x}}^{(s+1)} ,\boldsymbol{\rm{z}}^{(s+1)} \big]$. 

When the optical surfaces are sampled perpendicular to the $\hat{z}$ axis, the elements of the $z$ column vectors are identical. For example, for the optical system shown in Fig. \ref{fig:Lap_fig2}, the elements of $\boldsymbol{\rm{z}}^{(s)}$ are zero, while the elements of $\boldsymbol{\rm{z}}^{(s+1)}$ are equal to $z$. However, when inclined surfaces are considered, i.e. optical surfaces that are not sampled perpendicular to the $\hat{z}$ axis, the elements of the $z$ column vectors can be nonidentical, as will be discussed in Section \ref{sec:transmission_matrices}. Below, we will describe how propagation matrix $\boldsymbol{\rm{T}}^{(s)}$ is obtained for optical surfaces sampled perpendicular to the $\hat{z}$ axis, but this procedure is also applicable to optical system with inclined surfaces, as will be seen later.

The propagation matrix $\boldsymbol{\rm{T}}^{(s)}$, which has dimensions $M' \times N'$, describes the mapping of an electric field over surface $s$ to the surface $s+1$, i.e. the fields propagating to areas outside of the sampled surfaces are ignored. Here, we specifically choose to use a spherical-wave propagator to populate matrix $\boldsymbol{\rm{T}}^{(s)}$, but the numerical modal framework allows for any field propagation method to be used. In our case, the columns of matrix $\boldsymbol{\rm{T}}^{(s)}$, $\boldsymbol{\rm{t}}_{n'}^{(s)}$, are the complex column vectors containing the discrete electrical fields produced over surface $s+1$ when a point source is placed at the discrete positions over surface $s$, such that $\boldsymbol{\rm{T}}^{(s)} = \big[ \boldsymbol{\rm{t}}_{1}^{(s)}, \boldsymbol{\rm{t}}_{2}^{(s)},\dots,\boldsymbol{\rm{t}}_{N'}^{(s)} \big] $. 
The column vector elements of $\boldsymbol{\rm{t}}_{n'}^{(s)}$, i.e. $t_{m',n'}^{(s)}$, are obtained by applying the Huygens-Fresnel principle, which states that every point on a wavefront is a point source emitting a spherical wave: 
\begin{equation}
	t_{m',n'}^{(s)} = \big( z / \lambda \big)^{1/2} \text{exp} \big[ -i 2 \pi r_{m',n'}^{(s)} / \lambda + i \pi/4 \big] /r_{m',n'}^{(s)}.
	\label{eq:tranformation_matrix}
\end{equation}
This result can be obtained by evaluating the Fresnel diffraction integral in the limit of a point source and accounts for obliquity. Here, $r_{m',n'}^{(s)} = \Big\{ \big[z_{n'}^{(s)} - z_{m'}^{(s+1)}\big]^2 + \big[x_{n'}^{(s)} - x_{m'}^{(s+1)}\big]^2\Big\}^{1/2}$ is the position vector relating the $n'$-th position over the input surface to the $m'$-th position over the output surface, where $(x_{n'}^{(s)},z_{n'}^{(s)})$ and $(x_{m'}^{(s+1)},z_{m'}^{(s+1)})$ are the $x,z$ coordinates associated with the discrete positions over surface $s$ and surface $s+1$, respectively, and $\text{exp} (i \pi/4 ) $ is the Gouy phase factor \cite{Goldsmith:98}. Note that Eq. (\ref{eq:tranformation_matrix}) is the numerical equivalent of the free-space spherical-wave propagator and hence satisfies the wave-equation as well \cite{BornWolf:99,Wolf:07}. 
A graphical representation of the procedure for obtaining $\boldsymbol{\rm{T}}^{(s)}$ is given in Fig. \ref{fig:Lap_fig2}.

The spherical-wave propagator, as opposed to the commonly used plane-wave propagator, is crucial for two reasons. First, it allows for more generic surfaces shapes (e.g. inclined planes and aspheres) to be included, and potentially highly non-paraxial systems to be analysed. The HFMF can be generalized to include projection (obliquity) effects on the basis of the local surface normal. Second, it ensures that the principal diffraction effects are included, which is essential for accurately describing the behaviour of FIR systems in general.

In order to produce physically meaningful results, each propagation matrix $\boldsymbol{\rm{T}}^{(s)}$ must preserve power. Similar to the 1-D numerical adaptation of the spherical-wave propagator, the 1-D power normalization of $\boldsymbol{\rm{T}}^{(s)}$ is given by   
\begin{equation}
    \boldsymbol{\rm{\widetilde{T}}}^{(s)} = \sqrt{\boldsymbol{\Delta}^{(s+1)} } \boldsymbol{\rm{T}}^{(s)} \sqrt{\boldsymbol{\Delta}^{(s)}} ,
    \label{eq:normalization_transformation_matrix}
\end{equation}
where $\boldsymbol{\rm{\widetilde{T}}}^{(s)}$ is the normalized equivalent of $\boldsymbol{\rm{T}}^{(s)}$. In Eq. (\ref{eq:normalization_transformation_matrix}), $\boldsymbol{\Delta}^{(s)}$ and $\boldsymbol{\Delta}^{(s+1)}$ are diagonal matrices, and the multiplication order is again intentionally reversed to respect the ordering of the optical surfaces in accordance with the adopted matrix notation.
The sample step sizes of the discrete sample points are placed along the diagonal of $\boldsymbol{\Delta}^{(s)}$ and $\boldsymbol{\Delta}^{(s+1)}$:
\begin{align}
    \boldsymbol{\Delta}^{(s)}&= \text{diag} \Big\{ \big[ \Delta x_{1}^{(s)}, \Delta x_{2}^{(s)}, \dots, \Delta x_{N'}^{(s)} \big] \Big\} \text{ and }\\
    \boldsymbol{\Delta}^{(s+1)} &= \text{diag} \Big\{ \big[ \Delta x_{1}^{(s+1)}, \Delta x_{2}^{(s+1)}, \dots, \Delta x_{M'}^{(s+1)} \big] \Big\} .
\end{align}

The normalization in Eq. (\ref{eq:normalization_transformation_matrix}) is defined such that the HFMF can i) take into account the different sample step size, and ii) to ensure that the HFMF produces quantities with the units of power over the respective surfaces. This enables the use of more generic surface sampling schemes, such as the non-equidistant sampling of the surface, while ensuring power conservation. 

%%%%%%%%
\subsubsection{Transmission matrices}
\label{sec:transmission_matrices}

In the HFMF, the optical surfaces are defined such that they have an input and exit plane, both of which are sampled by the same number of discrete points. In the optically thin limit, the optical surfaces that have phase transforming properties, such as the lenses and the grating, are described by diagonal transmission matrices, while the apertures (i.e. optical surface without phase transforming properties) are described by unitary diagonal transmission matrices. From now on, we will omit the unitary matrices, because they leave the phase of the incident field unaltered and only slow down the numerical simulation.

In general, the phase transforming elements in a GS comprise the mirrors and the grating. When we apply the optically thin limit to the lenses, and they are reduced to thin lenses. The transmission matrix of a single, thin lens is a diagonal matrix
\begin{equation}
	\boldsymbol{\Theta}^{(L)} = \text{diag} \Big\{ \big[ \theta^{(L)}_{1}, \theta_1^{(L)}, \dots , \theta_{N_L}^{(L)} \big] \Big\},
	\label{eq:lens_approx1}
\end{equation}
where the diagonal elements are obtained using the thin-lens approximation \cite{Goldsmith:98}:
\begin{equation}
	\theta^{(L)}_i = \text{exp}\Big( \big\{ -i 2 \pi [x^{(L)}_i]^2 \big\}/ \big\{2\lambda f \big\} \Big) \text{ for } i = 1,2,\dots,N_L.
	\label{eq:lens_approx2}
\end{equation}
Here, $f$ is the focal length of the thin lens, and $^{(L)}$ is the thin-lens label, which follows the labels in Fig. \ref{fig:Lap_fig1}, i.e. $(L)=(L1),(L2),(L3)$. Moreover, $x^{(L)}_i$ is the $i$-th element of the discrete column vector, $\boldsymbol{\rm{x}}^{(L)} = \big[ x^{(L)}_1, x^{(L)}_{2}, \dots, x^{(L)}_{N_L} \big]^{T}$. The column vector $\boldsymbol{\rm{x}}^{(L)}$ describes the surface of the thin lens sampled at $N_L$ discrete positions parallel to $\hat{x}$. At these discrete positions, the phase transformation properties of the thin lens are applied to the incident electric field by multiplying the discretized column vector describing the field with the diagonal transmission matrix describing the thin lens.

The phase transforming properties of gratings are governed by the grating equation \cite{BornWolf:99},
\begin{equation}
	\sin[\alpha] + \sin[\beta(\lambda)] = u \lambda / d^\prime,
	\label{eq:grating_eq}
\end{equation}
where $\beta(\lambda)$ is the angle of reflectance, $u$ is order of interference, and $d^\prime$ is the groove period. The surface of the grating can be described by a diagonal matrix 
\begin{equation}
	\boldsymbol{\Theta}^{(G)}	= \text{diag} \Big\{ \big[ \theta^{(G)}_{1}, \theta^{(G)}_{2}, \dots , \theta^{(G)}_{N_G} \big] \Big\},
	\label{eq:grating_approx1}
\end{equation}
where
\begin{equation}
	\theta^{(G)}_j	= \text{exp} \big[ -i 2 \pi x^{(G)}_j \sin( \Delta \beta) / \lambda \big] \text{ for } j = 1,2,\dots,N_G .
	\label{eq:grating_approx2}
\end{equation}
This result can be obtained by rewriting Eq. (\ref{eq:grating_eq}) when considering that a transmission grating is a linear phase transformer, where we have used the definition $\Delta \beta = \beta(\lambda ) - \beta_0$. Moreover $^{(G)}$ labels the grating, and in Eq. (\ref{eq:grating_approx2}) $x^{(G)}_{j}$ is the $j$-th element of the discrete column vector, $\boldsymbol{\rm{x}}^{(G)} = \big[ x^{(G)}_1, x^{(G)}_2, \dots, x^{(G)}_{N_G} \big]^{T}$. The column vector $\boldsymbol{\rm{x}}^{(G)}$ contains $N_G$ discrete positions parallel to $\hat{x}$, which sample the surface of the grating. At these discrete positions, the linear phase transformation properties of the grating are applied to the incident electric field, as was described above for the thin-lens. Equation (\ref{eq:grating_approx2}) is an important result, because it describes how an incident field is dispersed by the grating with wavelength, which is its principal optical function. 

\subsubsection{Few-mode grating optics}
\label{sec:few_mode_grating_optics}

We now use the definitions of the transmission and propagation matrices in combination with Eq. (\ref{eq:theory_optics}), and apply them to the 1-D GS model shown in Fig. \ref{fig:Lap_fig1} to obtain the normalized system transformation matrix $\boldsymbol{\rm{\widetilde{H}}}$ of this GS. Starting from the input surface of the GS (the slit), moving through the optics, towards the output surface (the focal plane), we get
\begin{equation}
 \boldsymbol{\rm{\widetilde{H}}} = \boldsymbol{\rm{\widetilde{T}}}^{(7)} \boldsymbol{\Theta}^{(L3)} \boldsymbol{\rm{\widetilde{T}}}^{(6)}  \boldsymbol{\Theta}^{(G)} \boldsymbol{\rm{\widetilde{T}}}^{(5)}  \boldsymbol{\Theta}^{(L2)}  \boldsymbol{\rm{\widetilde{T}}}^{(4)}   \boldsymbol{\rm{\widetilde{T}}}^{(3)}  \boldsymbol{\Theta}^{(L1)}  \boldsymbol{\rm{\widetilde{T}}}^{(2)}  \boldsymbol{\rm{\widetilde{T}}}^{(1)}.
 \label{eq:theory_get_G1}
\end{equation}
The matrix $\boldsymbol{\rm{\widetilde{H}}}$ has dimensions $M \times N$, with $N$ and $M$ being the total number of sample points over the input and output surface of the GS, respectively, that describes the principal propagation characteristics of the system as a function of wavelength. In Eq.  (\ref{eq:theory_get_G1}), $\{ \boldsymbol{\rm{\widetilde{T}}}^{(s)} \, | s = 1,2,..., 7 \}$ are the normalized propagation matrices. For instance, $\boldsymbol{\rm{\widetilde{T}}}^{(1)}$ describes the propagation of the field over the input surface, i.e. the slit, to the second surface, i.e. $\text{FM1}$. In addition, $\boldsymbol{\Theta}^{(L1)}$, $\boldsymbol{\Theta}^{(L2)}$, and $\boldsymbol{\Theta}^{(L3)}$; and $\boldsymbol{\Theta}^{(G)}$, are the transmission matrices of the lenses and the grating, respectively, and the unitary diagonal transmission matrices of the apertures are excluded. 

At FIR wavelengths, the propagation of an electric field via an inclined surface can give rise to beam distortion effects \cite{Murphy:87}. Because of the inclination of the surface, different spatial regions over the wavefront of incident beam cover different distances upon propagation, causing the amplitude and phase of the electric field over the inclined surface to vary. In the GS scheme considered here, $\alpha$ is nonzero, while $\Delta \beta \neq 0$ when $\lambda \neq \lambda_0$, therefore, for most wavelengths, the front and back surface of the grating are inclined planes (see in Fig. \ref{fig:Lap_fig1}), and beam distortion effects are to be expected. 
The HFMF can account for these effects, because it relies on the spherical-wave instead of the plane-wave propagator. 
To explain this, we consider the collimator lens and the inclined front surface of the grating of the 1-D GS model. 
First, we assume that the column vectors containing the sampled $x,z$ coordinates of the collimator lens and the inclined grating front surface, and the corresponding sample step sizes of these discrete points, are known.
Next, we use the prescription given in Section \ref{sec:propagation_matrices} to calculate the normalized propagation matrix that describes the mapping from the collimator lens to the inclined front surface of the grating, i.e. $\boldsymbol{\rm{\widetilde{T}}}^{(5)}$ (see Eq. (\ref{eq:theory_get_G1})). In these calculations, Eq. (\ref{eq:tranformation_matrix}) ensures that the differences in propagation distance are accounted for by relying on the position vector that relates the sampled $x,z$ coordinates over the collimator lens and the inclined grating front surface. In other words, when an electric field is propagated from the collimator to the grating front surface, the beam distortion effects are incorporated because the spherical-wave propagator by definition accounts for the amplitude and phase variations of the electric field over the inclined surface.

For grating spectrometers in general, $|\alpha| \neq |\beta| $, and therefore these optical systems experience anamorphic magnification \cite{Schweizer:79}. This projection effect causes the beam diameter of the exiting beam to change with respect to the beam diameter of the beam incident as a function of wavelength. In the 1-D GS model considered here, this effect becomes particularly pronounced when $|\alpha|>>|\Delta \beta|$. The HFMF accounts for this effect as follows. First, we determine the sample step sizes of the discrete points over the front and the back surface of the grating, i.e. the elements of $\boldsymbol{\Delta}^{(G,f)}$ and $\boldsymbol{\Delta}^{(G,b)}$, where $^{(G,f)}$ and $^{(G,b)}$ label the front and back surface of the grating, respectively. Next, each element of $\boldsymbol{\Delta}^{(G,b)}$ is scaled in accordance with the anamorphic magnification factor, $\cos(\alpha)/\cos(\Delta \beta)$, while the elements of $\boldsymbol{\Delta}^{(G,f)}$ do not change, because $\alpha$ is constant. As a result, the effective beam diameter of the exiting beam varies with wavelength, while the beam diameter of the beam over the grating front surface is constant, and anamorphic magnification is accounted for.

Since the optical system is spatially finite, the information throughput of the system is limited. In this case, the linear system operator, here represented by the normalized system transformation matrix $\boldsymbol{\rm{\widetilde{H}}}$, is said to be a bounded Hilbert-Schmidt kernel. Consequently, we can decompose matrix $\boldsymbol{\rm{\widetilde{H}}}$ into a unique set of orthonormal basis vectors using the numerical linear algebra equivalent of the Hilbert-Schmidt decomposition \cite{Wolf:82,Withington:01,Withington:07}, i.e the SVD: 
\begin{equation}
	\boldsymbol{\rm{\widetilde{H}}} = \boldsymbol{\rm{U}} \boldsymbol{\Sigma} \boldsymbol{\rm{V}}^{\dagger}.
	\label{eq:theory_SVD}
\end{equation}
Here, $\boldsymbol{\rm{V}}$ is a unitary matrix of dimensions $N \times N$, with $N$ being the total number of discrete points over the input surface, and $\boldsymbol{\rm{U}}$ is a unitary matrix of dimensions $M \times M$, with $M$ being the total number of discrete points over the output surface. Furthermore, $\boldsymbol{\Sigma}$ is an $M \times N$ matrix containing the singular values of matrix $\boldsymbol{\rm{\widetilde{H}}}$, which are placed along its diagonal in decreasing order, while the rest of the remaining entries are zero. In our scheme, the singular values can attain a maximum value of unity, because matrix $\boldsymbol{\rm{\widetilde{H}}}$ is power normalized. 

Equation (\ref{eq:theory_SVD}) describes the optical system in terms of characteristic orthogonal vectors spanning the input plane (i.e. the columns of $\boldsymbol{\rm{V}}$) that map in a one-to-one correspondence to a set of characteristic orthogonal vectors over the output plane (i.e. columns of $\boldsymbol{\rm{U}}$) with a certain efficiency. We define the optical modes of the GS as the collection of individually coherent, but mutually fully incoherent, fields, that have nonzero singular values (or efficiencies). The GS optics is said to be few-mode when between 5-20 optical modes have an efficiency higher than a certain threshold value, which we arbitrarily set at $10\%$. 

The set of optical modes is unique for the optical system, and is primarily determined by the geometrical and optical parameters of the optical components constituting the system, and its operation wavelength. For example, the sampled spatial form of these optical modes, i.e. the columns of unitary matrices $\boldsymbol{\rm{V}}$ and $\boldsymbol{\rm{U}}$, and their efficiencies, will vary with wavelength, because matrix $\boldsymbol{\rm{\widetilde{H}}}$ varies with wavelength, as will be seen in Section \ref{sec:simulation_GM}. 

The optical modes map the partially coherent field $\boldsymbol{\rm{E}}$ from the entrance slit, through the optics, onto the focal plane \cite{Withington:01,Withington:07}
\begin{equation}
    \boldsymbol{\rm{E}}' = \boldsymbol{\rm{\widetilde{H}}}^{} \boldsymbol{\rm{E}} \boldsymbol{\rm{\widetilde{H}}}^{\dagger},
    \label{eq:theory_optics3}
\end{equation} 
where $\boldsymbol{\rm{E}}'$ represents the (spatial) correlation matrix at the focal plane. When combining Eq. (\ref{eq:theory_input2}), (\ref{eq:theory_SVD}), and (\ref{eq:theory_optics3}), we get that 
\begin{equation}
    \boldsymbol{\rm{E}}' = \boldsymbol{\rm{U}} \boldsymbol{\Sigma} \boldsymbol{\rm{V}}^{\dagger} \big< \boldsymbol{\rm{e}} \boldsymbol{\rm{e}}^{\dagger} \big> \boldsymbol{\rm{V}} \boldsymbol{\Sigma}^{\dagger} \boldsymbol{\rm{U}}^{\dagger}.
    \label{eq:theory_optics4}
\end{equation}
It is illustrative to select a single statistical instance from $\big< \boldsymbol{\rm{e}} \boldsymbol{\rm{e}}^{\dagger} \big>$, i.e. the electric field column vector $\boldsymbol{\rm{e}}$, which is single-mode, and see how it is propagated through the system. This process consist of three sequential operations. First, the electric field $\boldsymbol{\rm{e}}$ is mapped onto the unitary matrix $\boldsymbol{\rm{V}}$, resulting in a column vector consisting of decomposition coefficients. Second, the coefficients are scaled by their singular value. Third, and last, the result from the previous step is mapped in a one-to-one correspondence onto the unitary matrix $\boldsymbol{\rm{U}}$, and we get the statistical average of the electric field at the output plane $\boldsymbol{\rm{e}}'$. We can also backward propagate $\boldsymbol{\rm{e}}'$, i.e. from the focal plane to the slit, using $\boldsymbol{\rm{V}} \boldsymbol{\Sigma}^{\dagger} \boldsymbol{\rm{U}} \boldsymbol{\rm{e}}'$. 

The normalization of the propagation matrices allow for optical properties of electric field $\boldsymbol{\rm{e}}$ to be traced when it propagates through the GS. For example, electric field $\boldsymbol{\rm{e}}$ can be normalize such that i) it has unity power, and ii) to account for the sample step sizes of the discrete points. Both are important, because now  $\text{Tr}(\boldsymbol{\rm{E}})/\eta_0=1$, where $\text{Tr}$ denotes the trace, and $\eta_0$ is the impedance of free space. Next, we can use the HFMF to propagate correlation matrix $\boldsymbol{\rm{E}}$ to each of the optical surfaces, and the trace of the resulting correlation matrices provides the total power in the field over each optical surface. What we will see is that the total power in the electric field decreases as we move further into the optical system, due to truncation of the propagating field by the optical components.

In general, Eq. (\ref{eq:theory_optics3}) tells us that the correlation matrix $\boldsymbol{\rm{E}}^\prime$ varies with wavelength, because both the correlation matrix $\boldsymbol{\rm{E}}$ and the normalized system transformation matrix $\boldsymbol{\rm{\widetilde{H}}}$ vary with wavelength. However, the correlation matrix $\boldsymbol{\rm{E}}$ does not vary considerably with frequency, as discussed in Section \ref{sec:field_distribution_over_slit}. Therefore, only the calculation of matrix $\boldsymbol{\rm{\widetilde{H}}}$ has to be carried out for each discrete wavelength (or frequency) to obtain a polychromatic description of the system, and how it propagates correlation matrix $\boldsymbol{\rm{E}}$.

\subsubsection{Measurement of an input spectrum}
Until now, the emphasis was on correlation matrix $\boldsymbol{\rm{E}}$ and how it can be propagated to the output plane, without considering the spectral content explicitly. Normally, a grating spectrometer measures a spectral power per unit bandwidth, $\Delta \nu$. Here, we describe how the HFMF simulates the measurement of a partially coherent input spectrum.

First, we define a discrete frequency vector   
\begin{equation}
	\boldsymbol{\rm{\nu}} = \big[ \nu_1, \nu_2, \dots, \nu_{N_{\nu}} \big]^{T} ,
	\label{eq:input_frequency}
\end{equation}
that contains $N_{\nu}$ frequencies, each separated by a discrete step, $\Delta \nu_k$. From now on, $_k$ will be used to label the ${k}$-th discrete frequency, where $k = 1,2,\dots,N_{\nu} $. Next, we use $\boldsymbol{\nu}$ to define a discrete input spectral power over the slit,   
\begin{equation}
	\boldsymbol{\rm{b}} = \big[ b_1,b_2,\dots,b_{N_{\nu}} \big]^{T},
	\label{eq:b}
\end{equation} 
which has an arbitrary spectral form or Power Spectral Density (PSD). The elements of spectrum $\boldsymbol{\rm{b}}$ are in unit the power per unit bandwidth, and the quantity $b_k$ denotes the spectral power content at $\nu_k$. Then, for each ${k}$-th frequency, an correlation matrix describing the state of coherence of the incident spectral field over the slit, 
\begin{equation}
    \boldsymbol{\rm{B}} = b_k \boldsymbol{\rm{E}} ,
    \label{eq:B}
\end{equation}
is obtained. Finally, the spectral correlation matrix $\boldsymbol{\rm{B}}$ is propagated to the output plane according to Eq. (\ref{eq:theory_optics3}) 
\begin{equation}
    \boldsymbol{\rm{B}}' = \boldsymbol{\rm{\widetilde{H}}}^{} \boldsymbol{\rm{B}}^{} \boldsymbol{\rm{\widetilde{H}}}^{\dagger},
    \label{eq:B_prime}
\end{equation}
where $\boldsymbol{\rm{B}}'$ is the correlation matrix describing the state of coherence of the spectral field at the output plane at $\nu_k$.

\subsubsection{Straylight}
\label{sec:theory_straylight}
The performance of ultra-low noise space-based spectrometers can be greatly affected by straylight originating from regions surrounding the on-sky source, the instrument itself (i.e. internally generated emission), or by a combination of both. In this section, we will show how this radiation source is accounted for by the framework. Although on-sky straylight can affect the overall performance of a GS significantly \cite{Makiwa:13}, we will ignore it for now and we only focus on the effects of internally generated straylight. 

\begin{figure}[t!]
    \centering
    \includegraphics[width = 8.4 cm]{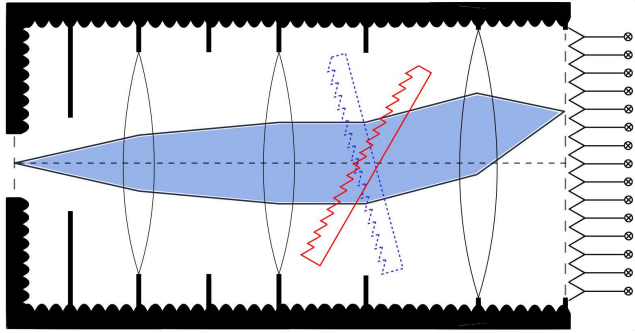}
    \caption{Grating spectrometer from Fig. \ref{fig:Lap_fig1} placed in an enclosure. The black patches represent the perfect electromagnetic absorber that covers the walls of the enclosure.}
    \label{fig:Lap_fig3}
\end{figure}

A FIR instrument is normally enclosed by a mechanical structure i) to support its optical components, and ii) to shield it from external radiation. We consider the enclosure to be a box of which the walls are covered with electromagnetic absorber (Fig. \ref{fig:Lap_fig3}). Here, we assume that the electromagnetic absorber is a perfect black body, such that all incident radiation onto the absorber, e.g. from reflections within the optical system, is absorbed. This perfect black body has a physical temperature, with typically $\,T_s < 4\,K$, and emits incoherent thermal background radiation. This straylight radiation is able to reach the detector plane, and therefore, in addition to spectral power from the slit, the GS also becomes sensitive to thermal emission originating from its mechanical and optical components. 

Adding this thermal straylight source to Eq. (\ref{eq:B_prime}) we obtain the total partially coherent radiation field at the focal plane for $\nu_k$, 
\begin{equation}
        \boldsymbol{\rm{Y}}' = \boldsymbol{\rm{\widetilde{H}}}^{} \boldsymbol{\rm{B}}^{} \boldsymbol{\rm{\widetilde{H}}}^{\dagger}  + \boldsymbol{\rm{C}}' , 
    \label{eq:E_prime_tot}
\end{equation}
where $\boldsymbol{\rm{C}}'$ is the correlation matrix describing the contribution due to straylight at the detector plane.  

We can derive a mathematical expression for straylight correlation matrix $\boldsymbol{\rm{C}}'$ when we assume that the electromagnetic absorber is i) held at a constant temperature, i.e. $\,T_s = $ constant, and ii) that it is in thermodynamic equilibrium with the components it encloses. Furthermore, the GS slit is covered by a slab of the electromagnetic absorber covering the walls, and it is held at $\,T_s$. In other words, the GS is enclosed by a perfect blackbody source. 

In this case, the point sources over the surface of the blackbody slab covering the slit are excited uniformly and incoherently, and the straylight field over this slab is fully incoherent. In addition, the GS acts as a blackbody cavity, and the straylight field over the detector plane is also fully incoherent. The correlation matrices describing these straylight fields for $\nu_k$ over the blackbody slab and the detector plane are identity matrices $\boldsymbol{\rm{I}}$ and $\boldsymbol{\rm{I}}'$ \cite{Kano:62,Withington:01}, and we get 
\begin{align}
    \boldsymbol{\rm{B}}   &= c_k \boldsymbol{\rm{I}} \label{eq:strayCorMatB} \text{ and } \\
    \boldsymbol{\rm{Y}}'  &= c_k \boldsymbol{\rm{I}}' \label{eq:strayCorMatF} .
\end{align}
Here, $\boldsymbol{\rm{I}}$ and $\boldsymbol{\rm{I}}'$ have dimensions $N \times N$ and $M \times M$, and the quantity $c_k$ denotes the spectral power content of the straylight at $\nu_k$, which follows the PSD of a blackbody
\begin{equation}
	c_k 	= \frac{h \nu_k}{\text{exp} \big[ (h \nu_k)/(k_b T_s) \big] - 1},
	\label{eq:DSF1}
\end{equation} 
where $h$ is Planck's constant and $k_b$ Boltzmann's constant.

Using Eq. (\ref{eq:strayCorMatB}) and (\ref{eq:strayCorMatF}), Eq. (\ref{eq:E_prime_tot}) transforms into
\begin{equation}
    c_k \boldsymbol{\rm{I}}' = c_k \boldsymbol{\rm{\widetilde{H}}}^{}  \boldsymbol{\rm{I}} \boldsymbol{\rm{\widetilde{H}}}^{\dagger}  + \boldsymbol{\rm{C}}',
\end{equation}
which combined with Eq. (\ref{eq:theory_SVD}) can be solved for straylight correlation matrix $\boldsymbol{\rm{C}}'$:
\begin{equation}
    \boldsymbol{\rm{C}}' = c_k \boldsymbol{\rm{U}} (\boldsymbol{\rm{I}}' - \boldsymbol{\Sigma}  \boldsymbol{\Sigma}^{\dagger} ) \boldsymbol{\rm{U}}^{\dagger},
    \label{eq:Sprime_2}
\end{equation}
where we used that $\boldsymbol{\rm{V}}$ and $\boldsymbol{\rm{U}}$ are unitary matrices. Moreover, matrix $\boldsymbol{\rm{\widetilde{H}}}$ is an Hilbert-Schmidt operator, therefore $\boldsymbol{\rm{\widetilde{H}}} \boldsymbol{\rm{\widetilde{H}}}^{\dagger}$ is Hermitian, the eigenvalues of which are real. Thus, $\boldsymbol{\Sigma} \boldsymbol{\Sigma}^{\dagger} = \boldsymbol{\Sigma}^2$, which are the power efficiencies, and Eq. (\ref{eq:Sprime_2}) becomes
\begin{equation}
    \boldsymbol{\rm{C}}' = c_k \boldsymbol{\rm{U}} (\boldsymbol{\rm{I}}' - \boldsymbol{\Sigma}^2 ) \boldsymbol{\rm{U}}^{\dagger}.
    \label{eq:Sprime_3}
\end{equation}
Here, the $\boldsymbol{\rm{U}} (\boldsymbol{\rm{I}}' - \boldsymbol{\Sigma}^2  ) \boldsymbol{\rm{U}}^{\dagger}$-term represents the straylight modes, which occur when the efficiencies of the optical modes are smaller than unity, or equivalently, when there are optical losses in the system. This naturally occurs for high-order optical modes, due to spatial filtering, as will be seen in Section \ref{sec:simulation_GM}. A different perspective is that the orthogonal set of vectors over the output surface of the GS (the columns of $\boldsymbol{\rm{U}}$) become less sensitive to the slit over input surface and have stronger coupling to internally generated radiation. This radiation is carried to the exit plane by the straylight modes.

Finally, we uncover the slit over the input surface, and using Eq. (\ref{eq:Sprime_3}) we rewrite Eq. (\ref{eq:E_prime_tot}) 
\begin{equation}
    \boldsymbol{\rm{Y}}' = b_k \boldsymbol{\rm{\widetilde{H}}}^{} \boldsymbol{\rm{E}} \boldsymbol{\rm{\widetilde{H}}}^{\dagger}  + c_k \boldsymbol{\rm{U}} (\boldsymbol{\rm{I}}' - \boldsymbol{\Sigma}^2 ) \boldsymbol{\rm{U}}^{\dagger}.
    \label{eq:E_prime_tot2}
\end{equation}
Here, correlation matrix $\boldsymbol{\rm{Y}}'$ describes the total partially coherent spectral field at the output plane as the sum of i) the spectral field originating from the slit, and ii) the internally generated straylight. 

\subsection{Detector array}
At the output plane of the GM a detector array, consisting of $N_d$ individual detectors, is placed to measure the power in the total incident partially coherent spectral field. The reception properties of the detectors in the array determine the coupling between the detector array and total incident partially coherent spectral field. For example, the functional form of the detector reception patterns are important, but in the case of few-mode detectors, the ratio between the pixel size and the wavelength must also be considered \cite{Chuss:08}. The HFMF is able to take into account these detector coupling  characteristics by using a correlation matrix that describes the proprieties of the detector array.  

In our simulations, we assume that the power over the aperture of the detector is equivalent to the power measured by the GS over that spectral bin. This simplifies the modelling, allowing us to focus on the coupled partially coherent behaviour, but it means that we ignore effects occurring inside the detector. Under these assumptions, we define a detector with label $i'$ to have a reception pattern
\begin{equation}
    \boldsymbol{\rm{D}}_{i'} = \sum_{k'}^{N_m} \sigma_{k'} \boldsymbol{\rm{d}}_{k'}^{} \boldsymbol{\rm{d}}_{k'}^{\dagger} \text{ for } i'=1,2,\dots,N_d ,
    \label{eq:theory_det2}
\end{equation}
where $\boldsymbol{\rm{d}}_{k'}$ is the electric field of the $k'$-th orthogonal detector mode over the aperture that has an efficiency $\sigma_{m'}$, with $k'= 1, 2, ..., N_m$ where $N_m$ is the total number of detector modes. 
We then define a detector array correlation matrix, $\boldsymbol{\rm{D}}$, which has the reception pattern of the detectors along its diagonal
\begin{equation}
    \boldsymbol{\rm{D}}= \left[ \begin{smallmatrix}
        \boldsymbol{\rm{D}}_1 &  &  &  \\
        & \boldsymbol{\rm{D}}_2 &  &  \\
        &  & \ddots &  \\
        & & & \boldsymbol{\rm{D}}_{N_d} 
        \end{smallmatrix} \right].
    \label{eq:theory_det1}
\end{equation}
The detector array correlation matrix $\boldsymbol{\rm{D}}$ captures the coupling characteristics of an individual detector, and those of the detector array as a whole. The main advantage of this approach is that, in principal, we can apply this formalism to any type of detector, such as KID \cite{Day:03} or TES \cite{Irwin:05} detectors, allowing us to construct any type of detector array. 

\subsection{Grating spectrometer description}

The total detected power at $\nu_k$ by the $i'$-th detector is given by $P_k^{i'}$. The quantity $P_k^{i'}$ is obtained by multiplying the spectral correlation matrix $\boldsymbol{\rm{Y}}'$ and the detector array correlation matrix $\boldsymbol{\rm{D}}$, taking the trace of this product over the aperture of the $i'$-th detector, and integrating it over the spectral bin $\Delta \nu_k$:
\begin{equation}
 	    P_{k}^{i'} = \text{Tr}^{i'} \big[ b_k \boldsymbol{\rm{D}}\boldsymbol{\rm{\widetilde{H}}}^{} \boldsymbol{\rm{E}} \boldsymbol{\rm{\widetilde{H}}}^{\dagger}  +  c_k \boldsymbol{\rm{D}}\boldsymbol{\rm{U}} (\boldsymbol{\rm{I}}' - \boldsymbol{\Sigma}^2 ) \boldsymbol{\rm{U}}^{\dagger} \big] \Delta\nu_k .
 	\label{eq:P_coupling}
\end{equation}
Here, $\text{Tr}^{i'}$ specifies that we are taking the trace over the aperture of the $i'$-th detector. 

Although Eq. (\ref{eq:P_coupling}) is useful, it only describes the detected power by a single detector at a single frequency; however, our goal is a polychromatic description. To obtain this description, we repeat Eq. (\ref{eq:P_coupling}) for each discrete frequency and each detector, such that 
\begin{equation}
    \boldsymbol{\rm{P}} = 
    \left[ \begin{smallmatrix}
        P^{1}_1 & P^{1}_2 & \cdots & P^{1}_{N_1-1} & P^{1}_{N_{\nu}}  \\
        P^{2}_{1} & P^{(2)}_{2} & \cdots & P^{2}_{N_{\nu}-1} & P^{2}_{N_{\nu}}  \\
        \vdots & \vdots & \vdots & \vdots & \vdots  \\
        P^{N_d-1}_{1} & P^{N_d-1}_{2} & \cdots & P^{N_d-1}_{N_{\nu}-1} & P^{N_d-1}_{N_{\nu}}  \\
        P^{N_d}_{1} & P^{N_d}_{2} & \cdots & P^{N_d}_{N_{\nu}-1} & P^{N_d}_{N_{\nu}} 
    \end{smallmatrix} \right] .
    \label{eq:power_detection_matrix}
\end{equation}
The measurement matrix $\boldsymbol{\rm{P}}$ has dimensions $N_d \times N_{\nu}$, the columns of which describe how the power in a discrete spectral input bin is distributed over the detector array, while its rows provide the spectral responses of each detector. In this process, the characteristics of i) the external and internal partially coherent fields, and ii) the few-mode nature of both the optics and the detector array, are taken into account simultaneously. These two effects have direct implications on the spectral response of the few-mode GS and the design of its components, which will become particularly pronounced when we analyze the performance of different GS designs in Section \ref{sec:results_system}. 

The measured spectrum is extracted from measurement matrix $\boldsymbol{\rm{P}}$ as follows. First, matrix $\boldsymbol{\rm{P}}$ is summed along its rows, resulting in the total detected power per detector. Then, using the GM geometry and Eq. (\ref{eq:grating_eq}), the centre of each detector aperture in the array is related to a specific frequency, and the measured spectrum is obtained by plotting the total detected power per detector as a function of this detector specific frequency. 

Measurement matrix $\boldsymbol{\rm{P}}$ also provides a metric for understanding how an incident spectral field, which can be in any state of coherence, is detected by a few-mode spectrometer, as a function of the geometrical parameters of the optical system, the detector geometry, straylight and wavelength. This metric could potentially be used as the basis for future spectral reconstruction techniques that will be required for analysing the performance of the next generation of ultra-low noise few-mode spectrometers.

An extensive set of simulations was conducted to verify the HFMF and to ensure that it reproduced known physical behaviour. The simulations included: i) standard Gaussian Beam Optics \cite{Goldsmith:98}; ii) reproducing Fresnel diffraction reported by \cite{Mahan:18}; iii) obtaining the optical modes of a pair of limiting 1-D apertures, which under certain conditions resulted in the Discrete Prolate Spheroidal Wavefunctions \cite{Slepian:61,Landau:61,Landau:62,Slepian:64,Slepian:78}; and iv) a 4f imaging system \cite{Goldsmith:98}, with a transmission grating placed at the intermediate Fourier plane. These simulations were used to validate the results obtained with the HFMF.

\section{Simulation results}
\label{sec:simulations}
We applied the HFMF to the SAFARI Long Wavelength Band as a representative case study to demonstrate its applicability and to show how it can be used for analyzing partially coherent FIR ultra-low noise systems. A few characteristic features are presented to illustrate the capabilities of the framework. 

The simulation results are divided into two parts: the GM optics and the grating spectrometer. In the first section, the few-mode characteristic of the GM optics are described and its response to two different inputs is analyzed. In the second section, the detector array is added, and together with the GM optics it forms the grating spectrometer. The performance of this GS system was investigated for two states of coherence over the input and two detector array types, using three spectral inputs. 

\subsection{Grating Module optics}
\label{sec:simulation_GM}
The Long Wavelength Band of SAFARI was designed to operate between $\lambda_{min}= 112\,\mu$m and $\lambda_{max} = 210\,\mu$m, and accepted an input beam with a focal ratio, $F = 5$. The physical and optical parameters of the grating module components are presented in Table \ref{tab:parameters_used1}, where $z$ is the distance to the next surface, $D$ is the aperture width, and $f$ is the focal length. The slit width, $a$, was determined by $a = A F \lambda_{max}$, where $A$ is an oversize factor of $1.5$. The widths listed in Table \ref{tab:parameters_used1} included this oversize factor. The low-resolution diffraction grating ($R\sim300$) was designed to have a groove period ($d^\prime$) of $0.184$\,$\mu$m and was operated in the first order of interference ($u=1$) under an angle of incidence ($\alpha$) of $50^\circ$. 

\begin{table}[b!]
    \centering
    \caption{\bf Physical and optical parameters of the SAFARI Grating Module optical elements }
        \begin{tabular}{m{1.2cm}m{1.2cm}m{1.2cm}m{1.2cm}}
            \hline
            Surface & $D$ (mm) & $z$ (mm) & $f$ (mm)  \\
            \hline
                Slit        & 1.58    & 30      & -   \\
            	FM1         & 8       & 40      & -   \\
            	L1          & 20      & 120     & 188 \\
        	    FM2         & 37.78   & 90      & -   \\
        	    L2          & 60      & 255     & 320 \\
            	Grating     & 90      & 315     & -   \\   
            	L3          & 240     & 350     & 350 \\
            \hline
        \end{tabular}
    \label{tab:parameters_used1}
\end{table}

Using the parameters presented in Table \ref{tab:parameters_used1} and Eq. (\ref{eq:theory_get_G1}), we obtained the normalized system transformation matrix $\boldsymbol{\rm{\widetilde{H}}}$ describing the GM optics for each discrete wavelength of interest. Here, we will use an equidistant sampling of $ \Delta x = \lambda/2 $ for all optical surfaces.

\begin{figure}[t!]
    \centering
    \includegraphics[width = 8.4 cm]{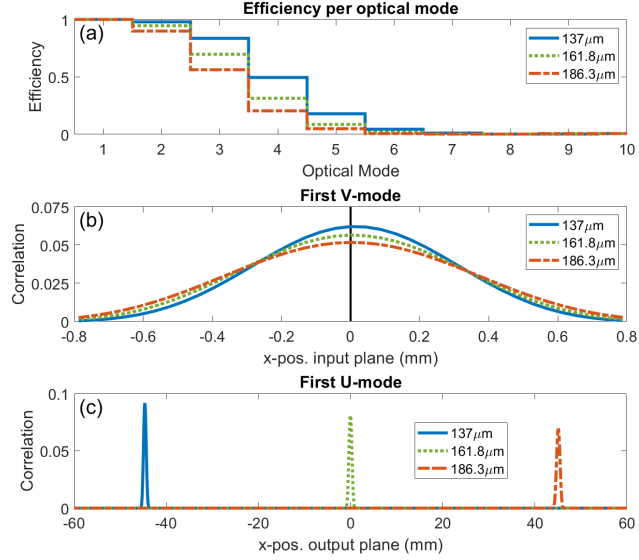}
    \caption{Few-mode behaviour of the GM optics as a function of wavelength. (a) gives the total number of optical modes and their efficiency, while (b) and (c) show the $\boldsymbol{\rm{V}}$ and $\boldsymbol{\rm{U}}$ component of the first optical mode.}
    \label{fig:Lap_fig4}
\end{figure}

Next, we determined the optical modes of the GM for three wavelengths: short ($\lambda_s = 137$\,$\mu\text{m'}$), centre ($\lambda_c = 161.8$\,$\mu\text{m'}$), and long ($\lambda_l = 186.3$\,$\mu\text{m'}$), where $\lambda_c$ is at the centre of the output plane, while $\lambda_s$ and $\lambda_l$ were chosen such that centre was located at $25\%$ from the exit slit edges. Then, using Eq. (\ref{eq:theory_SVD}) we determined the optical modes and their efficiencies. The first ten optical modes and their efficiencies, and the spatial form of the first optical mode over input and output plane (the first columns of unitary matrices $\boldsymbol{\rm{V}}$ and $\boldsymbol{\rm{U}}$) are shown in Fig. \ref{fig:Lap_fig4}. 

Four main observations are made looking at Fig. \ref{fig:Lap_fig4}. 
First, the total number of optical modes decreases with wavelength due to diffraction, and the GS is clearly few-mode. For the first optical mode, this translates into a change in the Full Width at Half Maximum of its spatial forms over the input and output plane as a function of wavelength. 
Second, the spatial form of the first mode over the input surface, i.e. the first column of unitary matrix $\boldsymbol{\rm{V}}$, is offset to the right (as seen from comparing the peak of the curve to the vertical black line at $x=0$ mm), due to the inclination of the grating front surface. 
Third, the spatial form of the first mode over the output surface, i.e. the first columns of unitary matrix $\boldsymbol{\rm{U}}$, shifts with wavelength. This behaviour is described by $\Delta \beta$ and it demonstrates that the HFMF accurately describes the dispersive properties of the GM optics. 
Fourth, the spatial forms of the first $\boldsymbol{\rm{V}}$- and $\boldsymbol{\rm{U}}$-mode follow a Gaussian-like profile, each with a Gaussicity $<99\%$. As mentioned previously, the slit was oversized with respect to the wavelength, and therefore the mapping between the input and output plane was accurately described by the fundamental (Gaussian) mode found in quasi-optical theory \cite{Goldsmith:98}. The close correspondence between the first optical mode and this fundamental quasi-optical mode confirmed that the HFMF provides physically meaningful results. 

Next, we focused on the correlations in the field, while ignoring the spectral content. To investigate how the spatial coherence of the input affected the performance of the GM optics, two inputs of the same spatial form, but in a different state of coherence, were selected. We chose the two extremes of coherence, i.e. the fully coherent and the fully incoherent case, because by simulating these limiting cases, we would be able to demonstrate that any input field, in any state of coherence, could be modelled by the framework.  

\begin{figure}[h!]
    \centering
    \includegraphics[width = 8.4cm ]{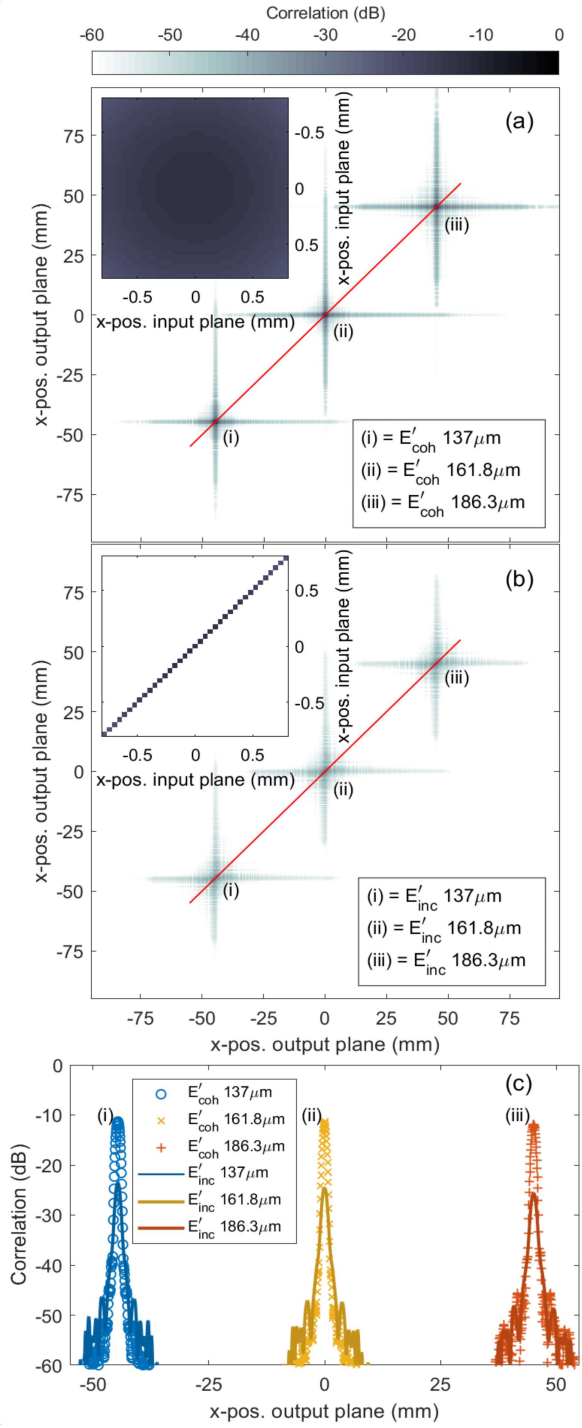}
    \caption{$\boldsymbol{\rm{E}}'$ as a function of input coherent state and wavelength. (a) and (b) show $\boldsymbol{\rm{E}}_{coh}'$ and $\boldsymbol{\rm{E}}_{inc}'$ for $\lambda_l, \lambda_c$ and $\lambda_u$, respectively. The insets show $\boldsymbol{\rm{E}}_{coh}$ and $\boldsymbol{\rm{E}}_{inc}$, and the red solid lines indicate the diagonals of $\boldsymbol{\rm{E}}_{coh}'$ and $\boldsymbol{\rm{E}}_{inc}'$, which are shown in (c).}
    \label{fig:Lap_fig5}
\end{figure}

The state of coherence of each case was described by a characteristic input correlation matrix $\boldsymbol{\rm{E}}$. From now on, $\boldsymbol{\rm{E}}_{coh}$ and $\boldsymbol{\rm{E}}_{inc}$ will denote the correlation matrix for the (fully) coherent and the (fully) incoherent input, which were obtained as follows. 
We defined $\boldsymbol{\mathsf{e}}$ to be a power normalized, discretely sampled, truncated Gaussian Beam \cite{Goldsmith:98}, 
\begin{equation}
    \boldsymbol{\mathsf{e}} = 
        \begin{cases} 
            \text{exp} \big\{ -[x_n^{(slit)}]^2 /a^2 \big\} & \text{for } |x_n^{(slit)}| \leq a/2 \\
            0       & \text{else where } ,
        \end{cases}
        \label{eq:assumed_field}
\end{equation}
for $n=1,2,\dots,N$, where $x_{n}^{(slit)}$ is the $n$-th element of the column vector $\boldsymbol{\rm{x}}^{(slit)}$ that contains the sample points over the slit, as defined in Eq. (\ref{eq:theory_input1}). Here, the slit width $a$ was chosen to be much smaller than the typical size of the diffraction pattern produced by the SPICA primary dish, such that the spatial distribution of the field over the slit could be decoupled from its spectral content. In addition, we were primarily interested in the behaviour of the grating optics, therefore we kept the Gaussian beam waist at the input fixed in order to keep the output beam waist fixed with respect to the detector.

Using Eq. (\ref{eq:theory_input2}) and Eq. (\ref{eq:assumed_field}), we obtained
\begin{equation}
    \boldsymbol{\rm{E}}_{coh} = \big< \boldsymbol{\mathsf{e}} \boldsymbol{\mathsf{e}}^\dagger \big>,
\end{equation}
while for the fully incoherent matrix   
\begin{equation}
    \boldsymbol{\rm{E}}_{inc} = \text{diag} \big( \big< \boldsymbol{\mathsf{e}}^2 \big> \big) ,
    %= \text{diag}\{[e^2(1),e^2(2),\dots, e^2(N_q)]^{T} \} 
\end{equation}
because only its diagonal elements were nonzero. The insets in Fig. \ref{fig:Lap_fig5}(a) and (b) show $\boldsymbol{\rm{E}}_{coh}$ and $\boldsymbol{\rm{E}}_{inc}$, respectively. These matrices were propagated through the GM optics using Eq. (\ref{eq:theory_optics3}), resulting in correlation matrices  $\boldsymbol{\rm{E}}_{coh}'$ and $\boldsymbol{\rm{E}}_{inc}'$ at the output plane, which are shown in Fig. \ref{fig:Lap_fig5} for $\lambda_l$, $\lambda_c$ and $\lambda_u$. Here, we observe again the dispersive properties of the grating, because the $x$-position corresponding with the maximum value in $\boldsymbol{\rm{E}}_{coh}'$ and $\boldsymbol{\rm{E}}_{inc}'$ shifts with wavelength. 

It is useful to look at the diagonals of correlation matrices $\boldsymbol{\rm{E}}_{coh}'$ and $\boldsymbol{\rm{E}}_{inc}'$ when examining the behaviour of the GM, because they contain the correlations that are directly related to the intensity of the field and the optical effects become most apparent along the diagonal. Figure \ref{fig:Lap_fig5}(c) shows the diagonal elements of $\boldsymbol{\rm{E}}_{coh}'$ and $\boldsymbol{\rm{E}}_{inc}'$ for $\lambda_l$, $\lambda_c$ and $\lambda_u$. Here, the non-normalized correlation strength between the spatial points in the field is shown, where the maximum correlation strength $= 1 = 0$ dB. 

From Fig. \ref{fig:Lap_fig5}(c) we see that for both cases the correlation strength decreases and that the correlations are more spread out over the output plane with wavelength, due to diffraction. However, the diffraction effects, such as the edge-ringing features \cite{Kintner:75}, are stronger for correlation matrix $\boldsymbol{\rm{E}}_{inc}'$. This was as expected, because for correlation matrix $\boldsymbol{\rm{E}}_{inc}$ each field point diffracted independently as a point source, resulting in more scattering than for the coherent case. Moreover, correlation matrix $\boldsymbol{\rm{E}}_{inc}$ was spatially filtered upon propagation, due to the finite size of the optical system, and off-diagonal elements were introduced in correlation matrix $\boldsymbol{\rm{E}}_{inc}'$. Thus, the correlation matrix $\boldsymbol{\rm{E}}_{inc}'$ was no longer fully incoherent, but it was transformed into a partially coherent field, due to diffraction of the optical components.

To gain a conceptual appreciation for the capabilities of the method, we now turn to the case where we included straylight at a single discrete wavelength, $\lambda$. Here, we simulated two cases: a closed ($a=0$) and an open slit ($a=AF\lambda_{max}$). For a closed slit, $\boldsymbol{\Sigma}$ was a null matrix and all the optical modes had zero efficiency, while for an open slit, Eq. (\ref{eq:theory_SVD}) was used to obtain $\boldsymbol{\Sigma}$ and a few optical modes had a nonzero efficiency (see Fig. \ref{fig:Lap_fig4}). Consequently, the straylight correlation matrix for a closed and an open slit at the focal plane, i.e. $\boldsymbol{\rm{C}}^{\prime}_{c}$ and $\boldsymbol{\rm{C}}^{\prime}_{o}$, where $_c$ and $_o$ are used to label a closed and an open slit, respectively, which were obtained using Eq. (\ref{eq:Sprime_3}) and omitting $c_k$, differed as well.

\begin{figure}[t!]
    \centering
    \includegraphics[width = 8.4 cm]{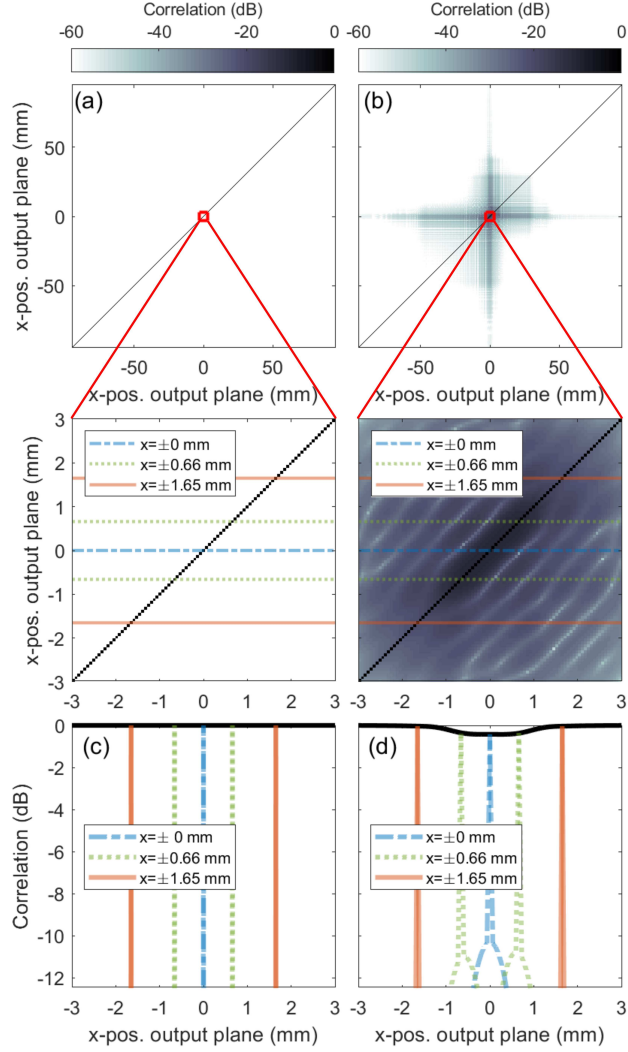}
    \caption{Straylight correlation matrix at the focal plane for $\lambda=\lambda_c$ for a closed ($\boldsymbol{\rm{C}}^{\prime}_{c}$) and an open slit ($\boldsymbol{\rm{C}}^{\prime}_{o}$) are shown in (a) and (b). The zoom in focuses on the area surrounding the slit, where the horizontal lines indicate the slicing positions used for obtaining the correlation functions. The latter are shown in (c) and (d) for $\boldsymbol{\rm{C}}^{\prime}_{c}$ and $\boldsymbol{\rm{C}}^{\prime}_{o}$, respectively. }
    \label{fig:Lap_fig6}
\end{figure}

We expected straylight correlation matrix $\boldsymbol{\rm{C}}^{\prime}_{c}$ to be diagonal, because the spatial points over the input plane were excited with equal amplitude and their relative phases were completely uncorrelated. On the other hand, for an open slit, the same amplitude excitation was used, but now the spatial points spanning the slit had a deterministic phase relationship, therefore straylight correlation matrix $\boldsymbol{\rm{C}}^{\prime}_{o}$ also had to contain nonzero off-diagonal elements. This was confirmed by Fig. \ref{fig:Lap_fig6}, which shows straylight correlation matrices $\boldsymbol{\rm{C}}^{\prime}_{c}$ and  $\boldsymbol{\rm{C}}^{\prime}_{o}$, and their corresponding correlation functions. The latter were obtained by slicing horizontally through matrices $\boldsymbol{\rm{C}}^{\prime}_{c}$ and $\boldsymbol{\rm{C}}^{\prime}_{o}$. In Fig. \ref{fig:Lap_fig6} we see that straylight correlation matrix $\boldsymbol{\rm{C}}^{\prime}_{c}$ only has diagonal entries and its correlation functions are $\delta$-functions, confirming that the spatial point over the input plane were (spatially) fully incoherent. However, correlations appear when the slit is uncovered, as seen from the off-diagonal elements, and the correlation functions of straylight correlation matrix $\boldsymbol{\rm{C}}^{\prime}_{o}$ are no longer $\delta$-functions. 

A physical intuitive explanation for this is given by looking at the set of basis functions ($\boldsymbol{\rm{U}}$-modes) that constitute straylight correlation matrices $\boldsymbol{\rm{C}}_{c}^{\prime}$ and $\boldsymbol{\rm{C}}_{o}^{\prime}$, when we transform from a closed to an open slit, and how they are related by a rotation matrix, $\boldsymbol{\rm{A}}$. When the slit is closed, all the spatial points over the output plane are spatially incoherent and orthonormal. In this case, rotation matrix $\boldsymbol{\rm{A}}$ can be used to rotate the elements of straylight correlation matrix $\boldsymbol{\rm{C}}^{\prime}_{c}$ into a basis in which each individual point over the output planes corresponds with a single mode (or $\delta$-function)
\begin{equation}
	\boldsymbol{\rm{C}}_{c}^{\prime\ast} = \boldsymbol{\rm{A}}^{\dagger} \boldsymbol{\rm{C}}_{c}^{\prime} \boldsymbol{\rm{A}} \text{ and } \boldsymbol{\rm{C}}_{c}^{\prime} = \boldsymbol{\rm{A}} \boldsymbol{\rm{C}}_{c}^{\prime\ast} \boldsymbol{\rm{A}}^{\dagger},
\end{equation}
where the $^{\ast}$ indicates that straylight correlation matrix $\boldsymbol{\rm{C}}^{\prime}_{c}$ is transformed by $\boldsymbol{\rm{A}}$.
Next, the slit is opened, allowing the electric field incident on the slit to enter the GM through a single, discrete point. As a result, the amplitude of the single mode corresponding to that point decreases
\begin{equation}
	\boldsymbol{\rm{C}}_{o}^{\prime\ast} = \boldsymbol{\rm{C}}_{c}^{\prime\ast} - \gamma \boldsymbol{\Delta}_{11},
\end{equation}
where $\gamma$ is the decrease in amplitude and $\boldsymbol{\Delta}_{11}$ is a null matrix except for the entry corresponding to the single, discrete point at the input, which is unity. In this basis, there are no correlations present, but if we rotate back we get
\begin{align}
\begin{split}
	\boldsymbol{\rm{C}}_{o}^{\prime} &= \boldsymbol{\rm{A}} \boldsymbol{\rm{C}}_{o}^{\prime\ast} \boldsymbol{\rm{A}}^{\dagger} \\
	&= \boldsymbol{\rm{A}} \boldsymbol{\rm{C}}_{c}^{\prime\ast} \boldsymbol{\rm{A}}^{\dagger} - \gamma \boldsymbol{\rm{A}} \boldsymbol{\Delta}_{11} \boldsymbol{\rm{A}}^{\dagger} \\
	&= \boldsymbol{\rm{C}}_{c}^{\prime} - \gamma (\boldsymbol{\rm{A}} \boldsymbol{\Delta}_{11} \boldsymbol{\rm{A}}^{\dagger}).
\end{split}
\end{align}
Here, straylight correlation matrix $\boldsymbol{\rm{C}}^{\prime}_{c}$ is diagonal, but correlation matrix $(\boldsymbol{\rm{A}} \boldsymbol{\Delta}_{11} \boldsymbol{\rm{A}}^{\dagger})$ is not. In other words, straylight correlation matrix $\boldsymbol{\rm{C}}^{\prime}_{o}$ is not diagonal and correlations appear between the spatial points over the output plane when the slit is opened.

These results showed that the method was able to reproduce and provide insight into various optical properties of the GM optics, such as the state of coherence of the propagating field, its few-mode wavelength-dependent behaviour, diffraction and straylight. The next step was to couple the detector array to the GM optics, and to investigate a few-mode FIR grating spectrometer.

\subsection{Grating Spectrometer}
\label{sec:results_system}

The few-mode SAFARI Long Wavelength Band was simulated by coupling the GM optics presented above to a detector array. First, we defined two detector array types and obtained the detector array correlation matrix $\boldsymbol{\rm{D}}$ for each. Subsequently, we defined four GS cases (or configurations) and three discrete input spectra to analyze the grating spectrometer. Here, the aim was i) to explore the behaviour of the method when applied to a 1-D GS, and ii) to gain a conceptual appreciation of the operation principles of few-mode grating spectrometers, including some of the relevant trade-offs for optimising its design. 

Table \ref{tab:parameters_used3} presents the parameters of the SAFARI Long Wavelength Band detector array that consisted of three subbands: the short-, centre- and long-wavelength, which are labeled by $h=s,c,l$, respectively, similarly to the three wavelengths used in section \ref{sec:simulation_GM}. Each feedhorn coupled detector had detector response function (DRF) in units $\text{WHz}^{-1}$, where $g$ was the gap between between two adjacent detector apertures. The subbands had a specific detector pitch, $p_{h} = d_{h} + g$, being the centre-to-centre distance between two detectors in a subband. Here, $d_h$ i the 1-D equivalent of the horn aperture, i.e. the horn entrance slit. Table \ref{tab:parameters_used3} also lists the total (linear) span of the detector array and the total number of detectors ($N_d$). Note that $g$ and $N_d$ were constant by design, while $p_{h}$ varied, to account for wavelength-dependent effects.

\begin{table}[b!]
    \centering
    \caption{\bf The SAFARI Long Wavelength Band detector specifications }
    \begin{tabular}{llll} 
        \hline
        	$p_s$ & 1.05 mm 	& $g$ 	        & 0.1 mm \\
        	$p_c$ & 1.294 mm 	& Total span    & 190 mm \\
        	$p_l$ & 1.594 mm	& $N_d$	        & 144 (48 per subband) \\
        \hline
    \end{tabular}
    \label{tab:parameters_used3}
\end{table}

In the simulations, the individual detectors were selected to be 1-D rectangular horn antennas to match the detectors of the SAFARI Long Wavelength Band. In this case, the $k'$-th detector aperture mode, $\boldsymbol{\rm{d}}_{k'}$, was described by
\begin{equation}
	\boldsymbol{\rm{d}}_{k'} = \big[ d^{(k')}_{1},d^{(k')}_{2},\dots,d^{(k')}_{N_a} \big]^{T},
	\label{eq:theory_det3}
\end{equation}
with
\begin{equation}
	d^{(k')}_r = \text{cos}\Big( j \pi x^{(a)}_{r}/d_h \Big) \, \text{for} \, r = 1,2,\dots, N_a .
	\label{eq:theory_det4}
\end{equation}
Here, $^{(k')}$ and $^{(a)}$ are used to label the detector aperture mode and the discrete sample points over the horn aperture, respectively. Furthermore, $x^{(a)}_{r}$ is the $r$-th element of $\boldsymbol{\rm{x}}^{(a)} = \big[ x^{(a)}_1, x^{(a)}_2, \dots, x^{(a)}_{N_a} \big]^{T}$, which is column vector containing the $N_a$ discrete $x$-positions sampling the aperture of each individual detector.

The number of detector modes ($N_m$) normally varies as a function of wavelength \cite{Makiwa:13} and using Eq. (\ref{eq:theory_det2}) it would be straightforward to define a few-mode detector. Here, however, these kind of detectors were ignored and only two limiting cases were considered: a single-mode ($N_m=1$) and a highly multi-mode ($N_m >> 1$) detector. In a later paper, we will report on how a few-mode, wavelength-dependent detector array affects the performance of few-mode FIR optical systems. 

\begin{figure}[t!]
    \centering
    \includegraphics[width = 8.4 cm]{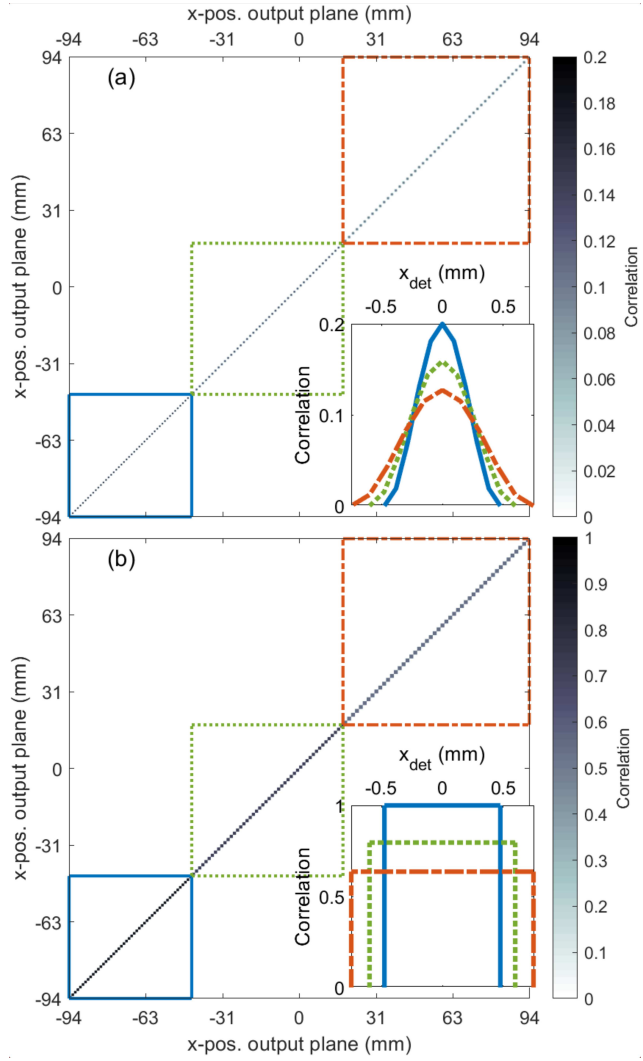}
    \caption{Detector array correlation matrix $\boldsymbol{\rm{D}}$ for the single-mode (a) and the highly multi-mode detector array (b). The subbands of the arrays are indicated by the dashed boxes: the left and upper right boxes represent the short and long subbands, respectively. The insets show the DRF for the short-, middle- and long-wavelength subband detectors indicated by the solid, dotted and dashed lines.}
    \label{fig:Lap_fig7}
\end{figure}

The single-mode detector (SMD) and highly multi-mode detector (MMD) array were each described by a detector correlation matrix (see Fig. \ref{fig:Lap_fig7}), which were obtained using Eq. (\ref{eq:theory_det2}), Eq. (\ref{eq:theory_det1}), Eq. (\ref{eq:theory_det3}) and Eq. (\ref{eq:theory_det4}). The DRFs of the SMD array were power normalized, first-order rectangular horn modes, and those of MMD array were boxcar functions. For the latter, the DRF of the short-wavelength subband had unity power over the aperture and the DRFs of the two other subbands were scaled accordingly to ensure power conservation.

Next, four GS configurations were defined based on the input correlations matrices defined in Section \ref{sec:theory} ($\boldsymbol{\rm{E}}_{coh}$ or $\boldsymbol{\rm{E}}_{inc}$) and the two detector array types (SMD or MMD). Furthermore, to investigate the GS design, its few-mode behaviour, and the performance of the four GS cases, we defined three discrete input spectra: $\boldsymbol{\rm{b}}_1$, $\boldsymbol{\rm{b}}_2$ and $\boldsymbol{\rm{b}}_3$. 
Their spectral forms were chosen such that they highlighted different aspects. For instance, $\boldsymbol{\rm{b}}_1$ provided the Point Spread Function (PSF), while $\boldsymbol{\rm{b}}_2$ was a generically representative astronomical spectrum, and $\boldsymbol{\rm{b}}_3$ was used to investigate straylight. All three input spectra were defined over the same frequency range and generated following the same procedure. 

The procedure for constructing a spectrum is best explain considering an arbitrary spectrum $\textbf{b}$, and the following two assumptions were made. First, we assumed that each discrete spectral element of $\textbf{b}$, i.e. $b_k$, followed the PSD of a blackbody (see Eq. (\ref{eq:DSF1})). In this case, the spectral form of $\textbf{b}$ can be defined using a discrete frequency-dependent temperature profile, $T_{\nu}$, and we obtained
\begin{equation}
	b_k = \frac{h \nu_k}{\text{exp}\big[ (h \nu_k)/(k_b T_{\nu_k} ) \big] - 1 } \text{ for } k = 1,2,\dots,N_{\nu} .
	\label{eq:PSD1_2}
\end{equation} 
Second, we assumed that an arbitrary astronomical spectrum could be constructed using three spectral features: i) broad-band continuum, ii) narrow unresolved line, and iii) broad resolved line, each described by a characteristic temperature profile. The broad-band continuum was represented by a thermal continuum background source at some constant physical temperature, $T_{con}$, such that its temperature profile was given by 
\begin{equation}
    T_{\nu_k,con} = T_{con} \quad \forall k \in N_{\nu}.
    \label{eq:T_freq_continiuum}
\end{equation}
The temperature profile of a narrow unresolved line, $T_{\nu,n}$, was characterized by a $\delta$-function centered at frequency, $\nu_n$, with a physical temperature, $T_n$, where the $_n$ label is now used to indicate that a narrow unresolved line is considered. In this case, 
\begin{equation}
    T_{\nu_k,n}  =
    \begin{cases} 
        T_n & \text{if } \nu_k = \nu_n \\
        0 & \text{elsewhere.}
    \end{cases}
    \label{eq:T_freq_narrow}
\end{equation}
The characteristic $T_{\nu_k}$ of a broad resolved line, $T_{\nu_k,b}$, was similar to that of the narrow line, but now the physical temperature, $T_b$, was drawn from a Gaussian distribution, $\mathcal{n'}(\nu_b,\sigma)$:
\begin{equation}
    T_{\nu_k,b}  =
    \begin{cases} 
        T_b \sim \mathcal{n'}(\nu_g,\sigma) & \text{if } \nu_g - \sigma \leq \nu_k \leq \nu_g + \sigma  \\
        0 & \text{elsewhere, }
    \end{cases}
    \label{eq:T_freq_broad}
\end{equation}
where $\nu_g$ is the centre frequency of the broad feature and $\sigma$ is the spectral width. In Eq. (\ref{eq:T_freq_narrow}) and Eq. (\ref{eq:T_freq_broad}), the labels $n$ and $b$ are used to indicate a narrow (unresolved) or broad (resolved) line, respectively. 

Next, using these two assumptions, spectrum $\textbf{b}$ was constructed, which consisted out of two steps. First, we defined the set of spectral features constituting $\textbf{b}$, which could contain any combination and multitude of the three spectral features. Each of these features had a physical temperature, and a centre frequency or frequency range, and we obtained their characteristic $T_{\nu_k}$'s using Eq. (\ref{eq:T_freq_continiuum})-Eq. (\ref{eq:T_freq_broad}). Second, we combined their individual $T_{\nu_k}$'s, resulting in a single $T_{\nu_k}$, which we used in Eq. (\ref{eq:PSD1_2}) to obtain spectrum $\textbf{b}$. 

\begin{table}[t!]
    \centering
    \caption{\bf Input spectra specifications in terms of wavelength }
    \begin{tabular}{llllll} 
        \hline
        Spectrum                    & Feature       & Type & $\lambda(\mu\text{m'})$  & $\sigma(\mu\text{m'})$ & $T$ (K) \\ \hline 
    	$\boldsymbol{\rm{b}}_1$     & $\delta(\lambda_n)$  & Emission      & $157.5$       & $\delta$-function  & 100 \\ 
        $\boldsymbol{\rm{b}}_2$     & Continuum     & Emission      & $[\lambda_{min},\ldots,\lambda_{max}]$ & -& 60\\
                                    & $\mathcal{n'}(\lambda_b,\sigma)$  & Absorption    & $131.8$       & $2.5$  & 56.4 \\
                                    & $\delta(\lambda_n)$       & Absorption    & $148.6$       & $\delta$-function    & 45  \\
                                    & $\delta(\lambda_n)$       & Emission      & $157.5$       & $\delta$-function    & 100 \\
                                    & $\mathcal{n'}(\lambda_b,\sigma)$  & Emission      & $181.8$       & $5$    & 63.7 \\
                                    & $\delta(\lambda_n)$       & Absorption    & $184.9$       & $\delta$-function    & 48.6  \\
    	$\boldsymbol{\rm{b}}_3$     & Continuum	    & Emission      & $[\lambda_{min},\ldots,\lambda_{max}]$ & - & 60 \\
                                    & $\delta(\lambda_n)$       & emission      & $157.5$	    & $\delta$-function    & 100 \\
        \hline 
    \end{tabular}
    \newline
    \label{tab:spectra}
\end{table}

The spectral characteristics of three discrete input spectra ($\boldsymbol{\rm{b}}_1$, $\boldsymbol{\rm{b}}_2$ and $\boldsymbol{\rm{b}}_3$) are listed in Table \ref{tab:spectra} in terms of wavelength. Both the narrow and broad line features were modelled in emission and absorption to mimic representative astronomical spectral features and to enable the reproduction of representative astronomical spectra. The input spectra were oversampled by a factor of eight with respect to $R$ ($\Delta \nu = \Delta \nu_k = 0.25$ GHz) to ensure that common, numerical artefacts, such as spectral aliasing, were avoided, Eq. (\ref{eq:theory_requirement}) was satisfied ($\bar{\nu}= 606$ GHz), and that the state of coherence of each spectrum was incorporated using Eq. (\ref{eq:B}).

The first input spectrum, $\boldsymbol{\rm{b}}_1$, was a narrow unresolved emission line (see Table \ref{tab:spectra}) and its measured spectrum (or spectral response) therefore provided the PSF for a given GS case. By studying this response we investigated the characteristics of the GM optics, the detector array, and the GS as a whole. This study consisted of two steps. First, the PSF for a single GS case was analyzed, to understand the distinct features in the PSF and their physical origin. Second, the PSFs of the four cases were compared in order to investigate how the GS properties changed with coherent state of the input and detector type.

\begin{figure}[h!]
\centering
\includegraphics[width = 8.4 cm]{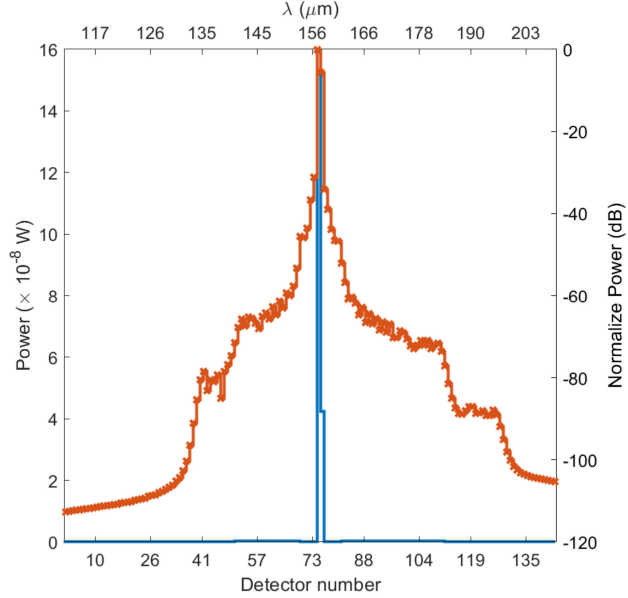}
\caption{Detection of the first spectrum ($\boldsymbol{\rm{b}}_1$) by the coherent/SMD GS configuration. The total spectral power is shown on a linear and dB scale by the blue and red line, respectively. For the latter, the spectral power is normalized to power detected by detector $\#48$.}
\label{fig:Lap_fig8}
\end{figure}

Figure \ref{fig:Lap_fig8} shows the PSF for the coherent/SMD case, which demonstrates that this GS configuration was able to reproduce the narrow feature. The spectral power falls primarily on two (peak) detectors, i.e. detector $\#75$ and $\#76$, due to the limiting spectral resolving power of the GS.  Moreover, three distinct features can be identified: i) the width of the recovered line; ii) the asymmetric profile of the spectral response; and iii) the edge-ringing effect that are typical of diffracting apertures \cite{Kintner:75}.

To investigate their physical origin, a select number of the optical parameters were changed, while keeping the waist of the input Gaussian beam fixed, and their PSFs were compared. From this analysis it became clear that the slit width ($a$) and incident angle of the grating ($\alpha$) were the driving parameters.
First, the width of the PSF was determined by the slit width, which, as expected, widened as the slit width increased in size. Second, the asymmetric profile of the PSF did not change with configuration and was intrinsic to the optics, or more specifically, to the grating. The grating was highly inclined and diffraction within its volume resulted in an asymmetric PSF profile at the output plane. 

\begin{figure}[t!]
\centering
\includegraphics[width = 8.4 cm]{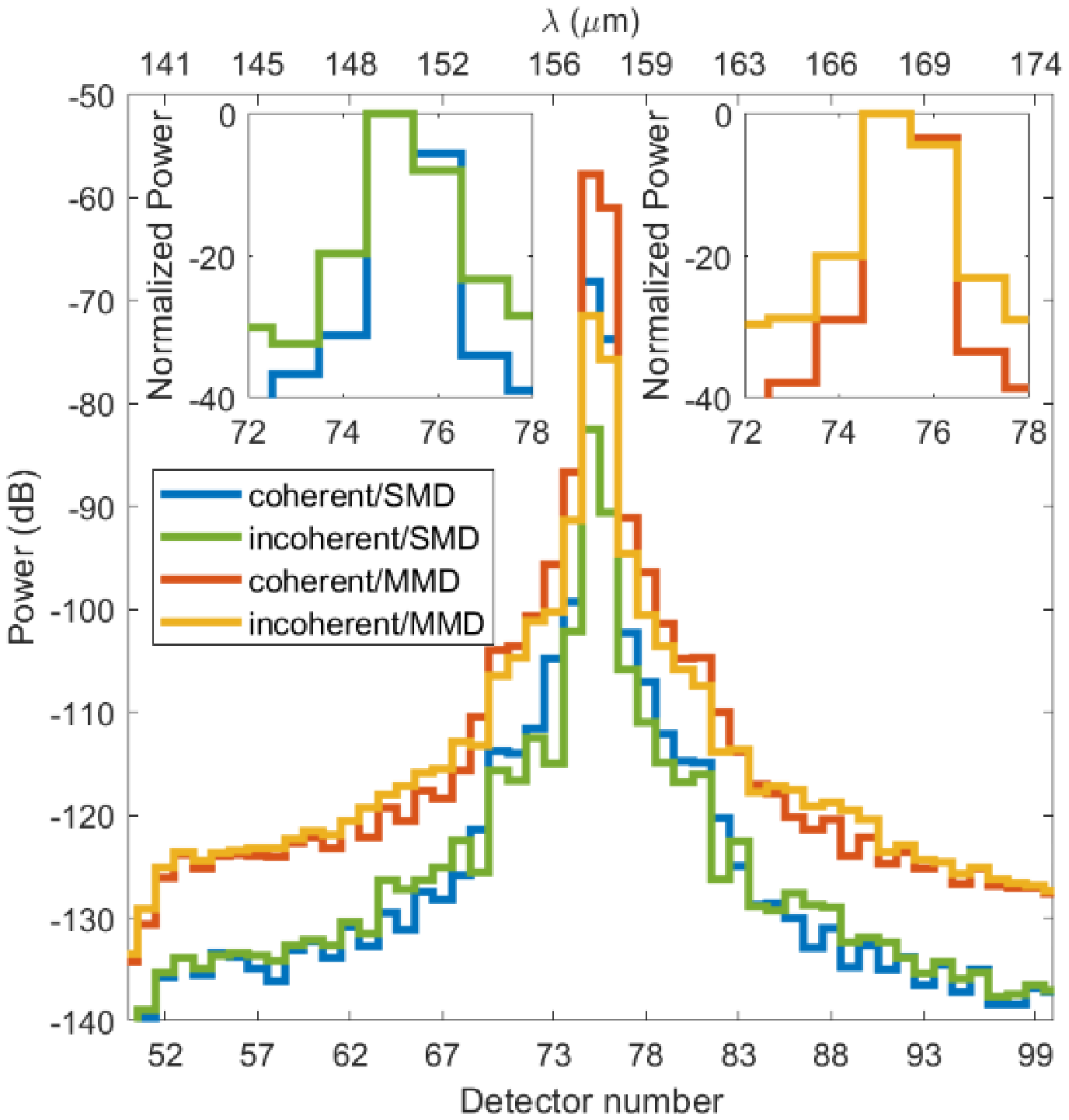}
\caption{PSF of the grating spectrometer as a function of the coherent state of the input (coherent or incoherent) and detector array type (SMD or MMD) for detector $\#50$ till $\#100$. The left and right inset provide a zoom on the normalized power of the peak and their neighboring detectors for the coherent/SMD and incoherent/SMD case, and the coherent/MMD and incoherent/MMD case, respectively. In these insets, the power is normalized to the power measured by detector $\#75$.}
\label{fig:Lap_fig9}
\end{figure}

In the second part of this analysis, we studied how the PSF changed with configuration, the results of which are shown in Fig. \ref{fig:Lap_fig9}. First of all, we see a difference in power of $\sim 10$ dB for all detectors when only the detector type is changed, for example, when comparing the lines in the upper left inset of Fig. \ref{fig:Lap_fig9}. On the other hand, when only the state of coherence over the slit changed, we see that the ratio of normalized power measured by the peak and the neighboring detectors changes. For instance, the insets of Fig. \ref{fig:Lap_fig9} show that the normalized power measured by the peaks detectors compared to the neighboring detectors is higher for the coherent case than for the incoherent case. 

The first observation was attributed to the fact that the MMD array contained more detector modes, resulting in a stronger overall spectral response of the GS (see the insets of Fig. \ref{fig:Lap_fig7}). The second observation was explained by the fact that the incoherent input experienced more diffraction upon propagation than the coherent input, because each field point diffracted independently as a point source, as discussed in Section \ref{sec:simulation_GM}. Therefore, in the case of an incoherent input, less of the input spectrum reached the detector plane, resulting predominantly in a decrease in the power measured by the peak detectors, because they were primarily sensitive to radiation originating from the input slit.

\begin{figure}[t!]
\centering
\includegraphics[width = 8.4 cm]{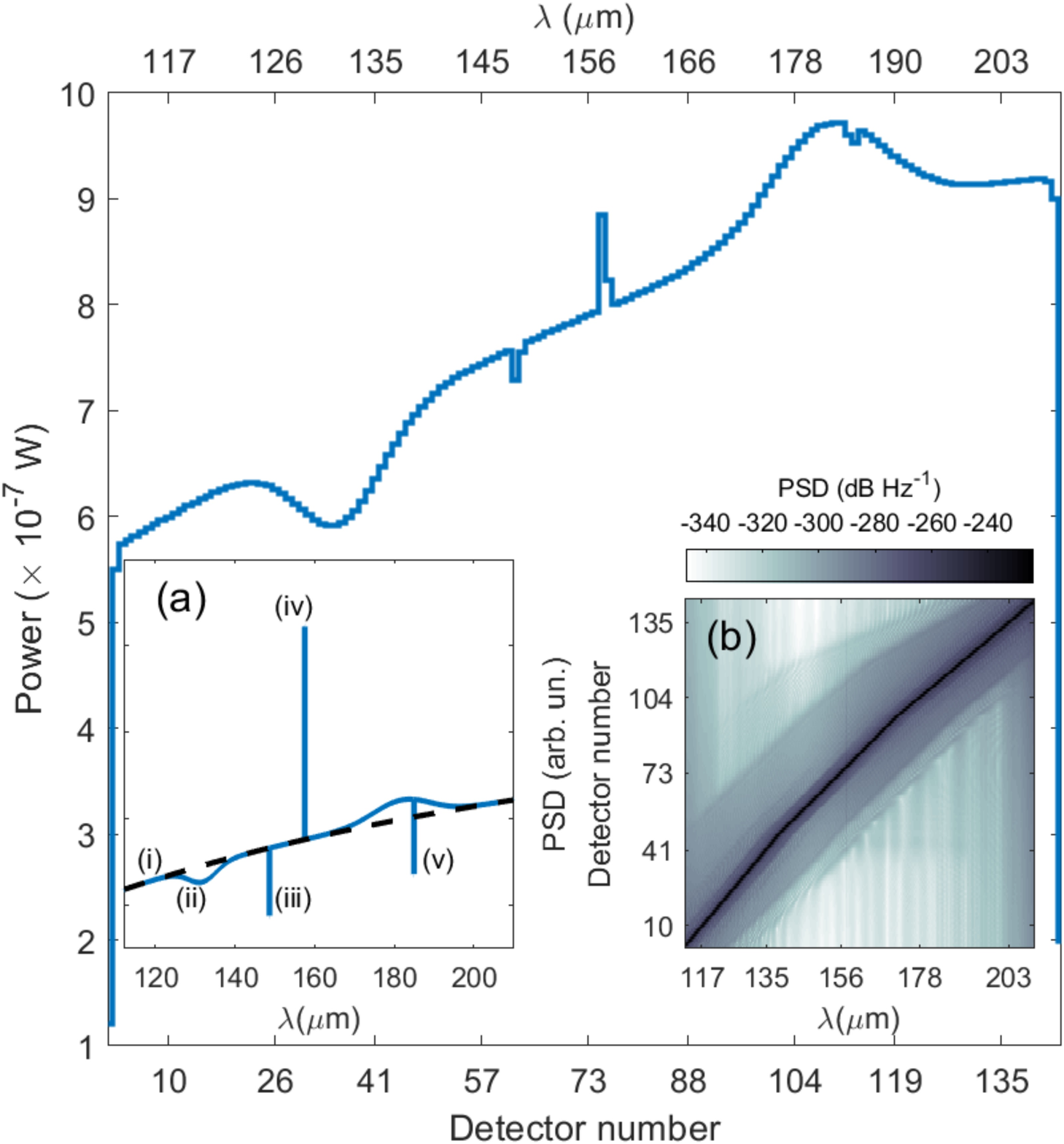}
\caption{Measured spectrum for the coherent/SMD case using the second input spectrum ($\boldsymbol{\rm{b}}_2$). The total power per detector is shown on a linear (the left axis) and a dB scale (the right axis). (a) shows the PSD, which consists of five representative spectral elements: i) a thermal background continuum indicated by the dashed black line; ii) a broad absorption feature; iii) and iv) a narrow absorption and emission line; and v) a multi-component feature, containing a narrow absorption line on top of a broadband emission. (b) shows the measurement matrix $\boldsymbol{\rm{P}}$ (see \ref{eq:power_detection_matrix}) for this case on a dB scale, which describes how $\boldsymbol{\rm{b}}_2$ is distributed over the detector array by the optics.}
\label{fig:Lap_fig10}
\end{figure}

To demonstrate that the measurement of more complex spectra could be simulated by the HFMF, we turned to the measurement of a representative astronomical spectrum $\boldsymbol{\rm{b}}_2$. Based on the result shown in Fig. \ref{fig:Lap_fig10}, we concluded that all the underlying spectral features in $\boldsymbol{\rm{b}}_2$ could be clearly recognized in the simulated spectrum, but the GS was unable to resolved the (unresolved) narrow lines due to its limiting resolving power. The details of the detection were captured by measurement matrix $\boldsymbol{\rm{P}}$ (see Eq. (\ref{eq:power_detection_matrix}) and Fig. \ref{fig:Lap_fig10}(b)), which described how the spectral content of $\boldsymbol{\rm{b}}_2$ was distributed over the detector array by the GM optics. Due to the grating, measurement matrix $\boldsymbol{\rm{P}}$ has a band structure, which relates a spectral bin, $\Delta \lambda_i$, to a specific detector (or $x$-position). 

\begin{figure}[t!]
\centering
\includegraphics[width= 8.4 cm]{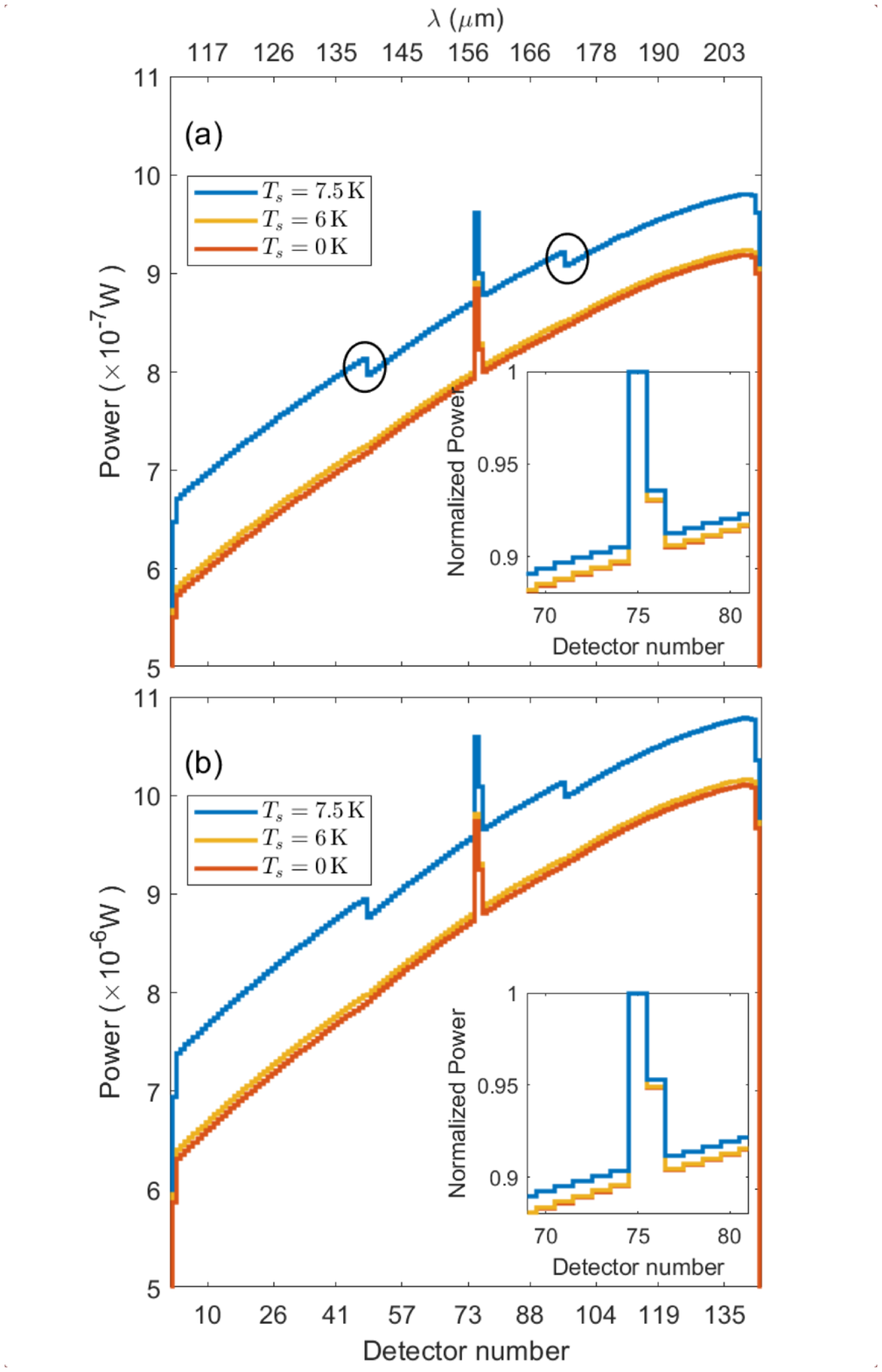}
\caption{Detection of the third spectrum ($\boldsymbol{\rm{b}}_3$) by the SAFARI Long Wavelength Band grating spectrometer as a function of straylight temperature ($T_s$) and detector array type using a fully coherent input correlation matrix. This corresponds with the GS measuring a spectrum from a point source imaged by some ideal fore optics. (a) and (b) show the power response for the coherent/SMD and coherent/MMD case, respectively. The features around detector $\#48$ and $\#96$ indicate the transitions between the subbands. The insets show a zoom in on the detector narrow feature, where the power is normalized to the detected power in detector $\#75$.}
\label{fig:Lap_fig11}
\end{figure}

\begin{figure}[t!]
\centering
\includegraphics[width= 8.4 cm]{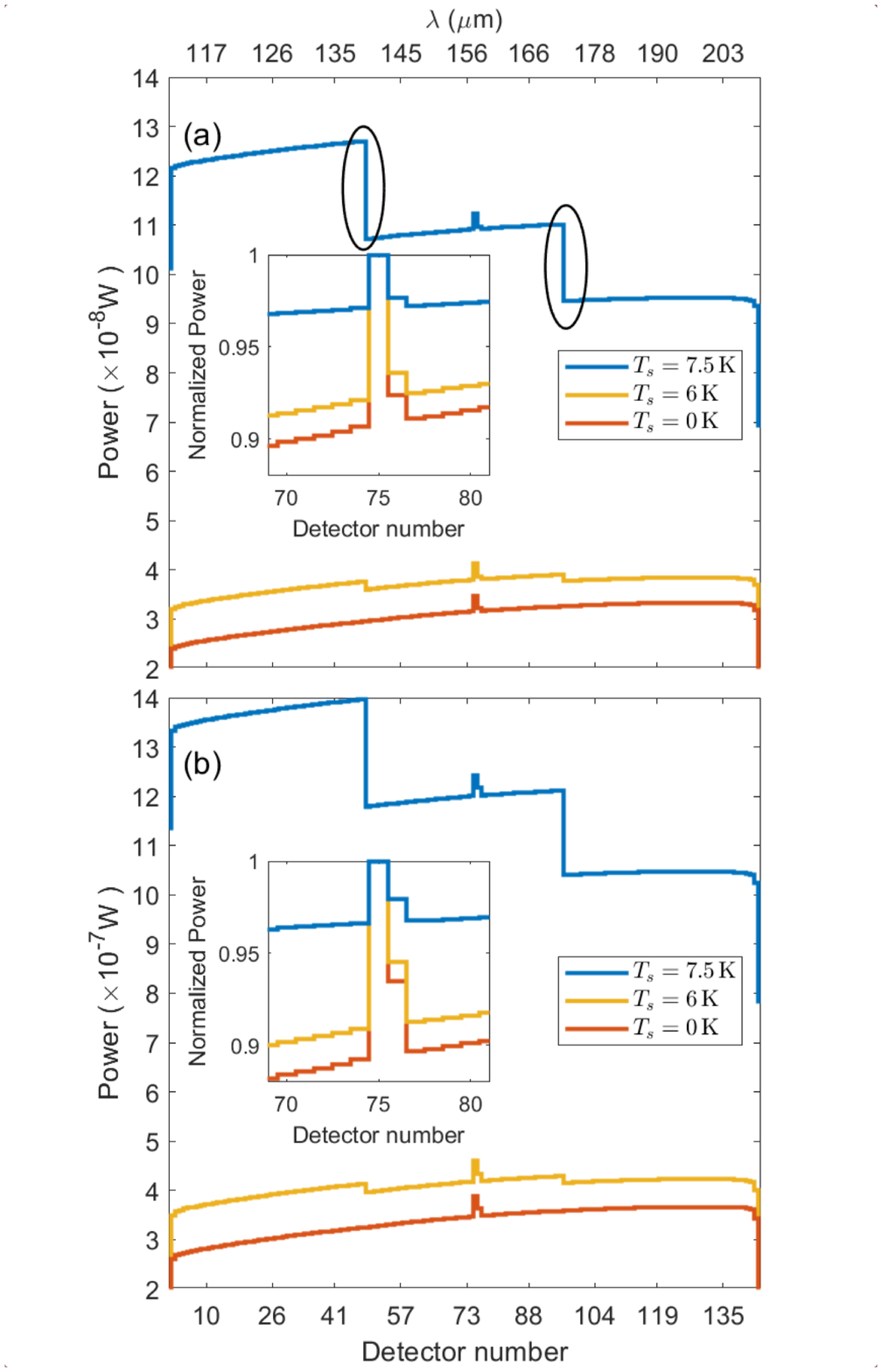}
\caption{Detection of the third spectrum ($\boldsymbol{\rm{b}}_3$) by the SAFARI Long Wavelength Band grating spectrometer as a function of straylight temperature ($T_s$) and detector array type using a fully coherent input correlation matrix. This corresponds with the GS measuring a spectrum from a blackbody source imaged by some ideal fore optics. (a) and (b) show the power response for the incoherent/SMD and incoherent/MMD case, respectively. The features around detector $\#48$ and $\#96$ indicate the transitions between the subbands. The insets show a zoom in on the detector narrow feature, where the power is normalized to the detected power in detector $\#75$.}
\label{fig:Lap_fig12}
\end{figure}

In the last set of simulations, straylight was included i) to demonstrate that the framework set out in this paper was capable of including this radiation source, and ii) to investigate the effects of straylight on the performance of the GS. For clarity we simplified spectrum $\boldsymbol{\rm{b}}_2$, such that the third spectrum ($\boldsymbol{\rm{b}}_3$) consisted only of the thermal background continuum and a narrow emission line.
Using Eq. (\ref{eq:DSF1}) and Eq. (\ref{eq:Sprime_3}), the contribution from internally generated straylight was determined. In the simulations, the straylight temperature ($T_s$) was varied to simulate a i) zero ($T_s=0$\,K), ii) weak ($T_s=6$\,K) and iii) strong ($T_s=7.5$\,K) straylight environment. The simulation results are shown in Fig. \ref{fig:Lap_fig11} and Fig. \ref{fig:Lap_fig12}, for the coherent and incoherent case, respectively. 

From Fig. \ref{fig:Lap_fig11} and \ref{fig:Lap_fig12} three main observations can be made. First, by looking at the insets we can see that the line-to-continuum changes with increasing $T_s$. Second, the transitions between the subbands (or subband structure) occurring around detectors $\#48$ and $\#96$ become apparent, as indicated by the black circles and ellipses in Fig. \ref{fig:Lap_fig11}(a) and \ref{fig:Lap_fig12}(a), respectively. Furthermore, this subband structure becomes more dominant for stronger straylight environments, which is particularly pronounced in Fig. \ref{fig:Lap_fig12}. Third, the global relative spectral response decreases with wavelength. 

Both the change in the line-to-continuum (see insets of Fig. \ref{fig:Lap_fig11} and \ref{fig:Lap_fig12}) and the appearance of the subband structures can be explained by how the spectral and straylight fields, i.e. correlation matrix $\boldsymbol{\rm{E}}^{\prime}$ and straylight correlation matrix $\boldsymbol{\rm{C}}^{\prime}$, couple differently to the detector array. The spatial form of correlation matrix $\boldsymbol{\rm{E}}^{\prime}$ is Gaussian-like and it shifts across the focal plane with wavelength (see Fig. \ref{fig:Lap_fig5}(c)), while straylight correlation matrix $\boldsymbol{\rm{C}}^{\prime}$, which is described by the Planck function, varies slowly with wavelength for the defined $T_s$ (see Fig. \ref{fig:Lap_fig6}(b)). 
In other words, the detector array measures the input spectrum on-top of a thermal background, which is constant, both in a spatial and spectral sense, and this causes the line-to-continuum to change. Furthermore, this constant background illumination also exposes subband structures, due to its constant nature. For instance, the subband structures were not seen for correlation matrix $\boldsymbol{\rm{E}}^{\prime}$, because matrix $\boldsymbol{\rm{E}}^{\prime}$ averages out the subband structures as its shifts spatially across the output plane with wavelength. However, straylight correlation matrix $\boldsymbol{\rm{C}}^{\prime}$ is unable to do so, because it is constant with wavelength. As a result, the DRF of the detector array is imposed onto the measured spectrum, and this effect becomes even more dominant for stronger straylight environments.  

The decrease in the global relative spectral response with wavelength for stronger straylight environment can be attributed to two effects. First, the straylight field is only constant by approximation, because in reality it decreases with increasing wavelength. Therefore, the straylight contribution is less for longer wavelengths, resulting in a lower relative spectral response than for shorter wavelengths. Second, intrinsic diffraction effects of the instrument are more dominant for longer wavelengths, which causes the global relative spectral response to decrease with wavelength.

\section{Conclusion}
\label{sec:dis}
We described the partially coherent modelling of few-mode FIR grating spectrometers.
The modal framework used for the simulations i) enables the propagation of the partially coherent fields, and ii) includes straylight coming from internal thermally radiating surfaces. This method enables the spatial-spectral performance of complex FIR optical systems to be determined within a single theoretical framework. Here, we focused on using the modal framework in combination with the Huygens-Fresnel principle, which together formed the HFMF, to demonstrate the partially coherent modelling of few-mode FIR grating spectrometers, where we used the grating spectrometer proposed for SPICA/SAFARI Long Wave Band as a case-study. 

First, we used the HFMF to analyze the behaviour of the GM optics without the detector array to illustrate its few-mode behaviour as a function of coherent state of the input and wavelength. The HFMF involves populating a normalized system transformation matrix $\boldsymbol{\rm{\widetilde{H}}}$ by applying the Huygens-Fresnel principle as the numerical equivalent of the free-space Green’s function. This matrix is central for two reasons: i) it allows for any spectral input field, in any state of coherence, to be incorporated, and ii) its SVD provides the optical modes of the GM, used for propagating the partially coherent input field. 

A detector array has subsequently been coupled to the GM optics, and four GS cases were defined based on the extreme states of coherence of the input, i.e. fully coherent and fully incoherent, and two detector array types. The performance of these cases was analyzed using three different input spectra. In the first part of this analysis, we investigated the physical concepts underlying a few-mode GS, and we explained how the HFMF can be used in GS design and performance analyses, e.g. when scaling optical components or downselecting the detector array design. In the second part, straylight was included to demonstrate how it affected the performance of few-mode FIR grating spectrometers. These results demonstrated that for the design of ultra-low-noise FIR spectrometers it is essential to have a rigorous understanding of i) the state of coherence of the source; ii) the few-mode behaviour of the detectors; and iii) the coupling mechanisms and characteristics of internally generated straylight radiation, since each affects the performance of the GS differently. 

Based on these results, we can conclude that the modal framework accurately describes the diffractive, dispersive, and few-mode characteristics of FIR optical system, and effectively handles various important matters for few-mode FIR ultra-low noise grating spectrometers in specific. For instance in the presence of straylight, where the HFMF detected features in the measured spectrum that would have been missed otherwise. Understanding these instrument characteristics with a high degree of confidence is crucial for future FIR spectroscopic missions, and demonstrates the utility of the framework in identifying the data reduction and calibration challenges that will be posed by the next generation of FIR space borne astronomical spectrometers. 

The HFMF is well equipped to serve multiple analyses and design purposes. From a physical optics point of view, it can be used to investigate the complex behaviour of ultra-sensitive FIR systems, such as diffraction and dispersion effects, as well as their modal behaviour. In the longer term, the HFMF could be used as an instrument simulator for FIR astronomy to investigate how astronomical metrics, e.g. individual line features, total-line-fluxes or line-to-continuum ratios, are affected by the partially coherent properties of the source, optics and the detector array.
From a design perspective, there is also wide range of applications. The span from focused projects, such as investigating the implications of partially coherent behaviour on the detector array design and instrument calibration, to the partially coherent analysis of a space missions using few-mode spectrometers, e.g. OST \cite{Leisawitz:18}. 

The HFMF is not limited to grating spectrometers only, and the general formulation presented in this paper can easily be expanded to more complex few-mode FIR spectrometers, such as an FTS and PDFTS. These systems can achieve unprecedented sensitivity and high spectral resolution, but only when the background loading and the photon noise is kept to a minimum. In a future paper, we will report how the HFMF can be used to investigate issues related to straylight in these types of broadband FIR spectrometers, how this technique can be utilized when developing spectral and spatial calibration strategies to mitigate these challenges, and how the framework can be extended to include polarization, to enable to partially coherent modelling of Post Dispersed Polarizing Fourier Transform Spectrometers.

\section*{Acknowledgements}
The SAFARI project in the Netherlands is financially supported through NWO grant for Large Scale Scientific Infrastructure nr 184.032.209. From the Canadian side, there have been financial support from the Canadian Space agency, the Canada foundation for Innovation and NSERC.

\textbf{Disclosures.} The authors declare no conflicts of interest.

\textbf{Data availability.} Data underlying the results presented in this paper are not publicly available at this time but may be obtained from the authors upon reasonable request.

% Bibliography
% \bibliographystyle{osajnl}
\bibliography{references}

\begin{thebibliography}{10}
\newcommand{\enquote}[1]{``#1''}

\bibitem{Jackson11}
B.~D. Jackson, P.~A.~J. de~Korte, J.~van~der Kuur, P.~D. Mauskopf, J.~Beyer,
  M.~P. Bruijn, A.~Cros, J.-R. Gao, D.~Griffin, R.~den Hartog, M.~Kiviranta,
  G.~de~Lange, B.-J. van Leeuwen, C.~Macculi, L.~Ravera, N.~Trappe, H.~van
  Weers, and S.~Withington, \enquote{The spica-safari detector system: Tes
  detector arrays with frequency-division multiplexed squid readout,}
  {\protect\JournalTitle{IEEE Transactions on Terahertz Science and
  Technology}} \textbf{2}, 12--21 (2012).

\bibitem{Roelfsema:2018}
P.~R. Roelfsema, H.~Shibai, L.~Armus, D.~Arrazola, M.~Audard, M.~D. Audley,
  C.~Bradford, I.~Charles, P.~Dieleman, Y.~Doi, and et~al., \enquote{Spica—a
  large cryogenic infrared space telescope: Unveiling the obscured universe,}
  {\protect\JournalTitle{Publications of the Astronomical Society of
  Australia}} \textbf{35} (2018).

\bibitem{Leisawitz:18}
D.~Leisawitz, E.~Amatucci, R.~Carter, M.~DiPirro, A.~Flores, J.~Staguhn, C.~Wu,
  L.~Allen, J.~Arenberg, L.~Armus, C.~Battersby, J.~Bauer, R.~Bell, P.~Beltran,
  D.~Benford, E.~Bergin, C.~M. Bradford, D.~Bradley, D.~Burgarella, S.~Carey,
  D.~Chi, A.~Cooray, J.~Corsetti, E.~D. Beck, K.~Denis, L.~Dewell, M.~East,
  S.~Edgington, K.~Ennico, L.~Fantano, G.~Feller, D.~Folta, J.~Fortney,
  J.~Generie, M.~Gerin, Z.~Granger, G.~Harpole, K.~Harvey, F.~Helmich,
  L.~Hilliard, J.~Howard, M.~Jacoby, A.~Jamil, T.~Kataria, S.~Knight,
  P.~Knollenberg, P.~Lightsey, S.~Lipscy, E.~Mamajek, G.~Martins, M.~Meixner,
  G.~Melnick, S.~Milam, T.~Mooney, S.~H. Moseley, D.~Narayanan, S.~Neff,
  T.~Nguyen, A.~Nordt, J.~Olson, D.~Padgett, M.~Petach, S.~Petro, J.~Pohner,
  K.~Pontoppidan, A.~Pope, D.~Ramspacher, T.~Roellig, I.~Sakon, C.~Sandin,
  K.~Sandstrom, D.~Scott, K.~Sheth, J.~Steeves, K.~Stevenson, L.~Stokowski,
  E.~Stoneking, K.~Su, K.~Tajdaran, S.~Tompkins, J.~Vieira, C.~Webster,
  M.~Wiedner, E.~L. Wright, and J.~Zmuidzinas, \enquote{The origins space
  telescope: Mission concept overview,}  (2018), Proc. SPIE 10698. Space
  Telescopes and Instrumentation 2018: Optical, Infrared, and Millimeter Wave,
  1069815 (24 July 2018).

\bibitem{Duncan:19}
D.~Farrah, K.~E. Smith, D.~Ardila, C.~M. Bradford, M.~J. DiPirro,
  C.~Ferkinhoff, J.~Glenn, P.~F. Goldsmith, D.~T. Leisawitz, T.~Nikola,
  N.~Rangwala, S.~A. Rinehart, J.~G. Staguhn, M.~Zemcov, J.~Zmuidzinas,
  J.~Bartlett, S.~J. Carey, W.~J. Fischer, J.~R. Kamenetzky, J.~Kartaltepe,
  M.~D. Lacy, D.~C. Lis, L.~S. Locke, E.~Lopez-Rodriguez, M.~MacGregor,
  E.~Mills, S.~H. Moseley, E.~J. Murphy, A.~Rhodes, M.~J. Richter,
  D.~Rigopoulou, D.~B. Sanders, R.~Sankrit, G.~Savini, J.-D. Smith, and
  S.~Stierwalt, \enquote{{Review: far-infrared instrumentation and
  technological development for the next decade},}
  {\protect\JournalTitle{Journal of Astronomical Telescopes, Instruments, and
  Systems}} \textbf{5}, 1 -- 34 (2019).

\bibitem{kamp:21}
I.~Kamp, M.~Honda, H.~Nomura, M.~Audard, D.~Fedele, L.~B. F.~M. Waters,
  Y.~Aikawa, A.~Banzatti, J.~E. Bowey, M.~Bradford, C.~Dominik, K.~Furuya,
  E.~Habart, D.~Ishihara, D.~Johnstone, G.~Kennedy, M.~Kim, Q.~Kral, S.~P. Lai,
  B.~Larsson, M.~McClure, A.~Miotello, M.~Momose, T.~Nakagawa, D.~Naylor,
  B.~Nisini, S.~Notsu, T.~Onaka, E.~Pantin, L.~Podio, P.~R. Marichalar,
  W.~R.~M. Rocha, P.~Roelfsema, F.~Santos, T.~Shimonishi, Y.~W. Tang,
  M.~Takami, R.~Tazaki, S.~Wolf, M.~Wyatt, and N.~Ysard, \enquote{The formation
  of planetary systems with spica,}  (2021).

\bibitem{Wiedemann:89}
G.~Wiedemann, D.~E. Jennings, R.~H. Hanel, V.~G. Kunde, S.~H. Moseley, G.~Lamb,
  M.~D. Petroff, and M.~G. Stapelbroek, \enquote{Postdispersion system for
  astronomical observations with fourier transform spectrometers in the thermal
  infrared,} {\protect\JournalTitle{Appl. Opt.}} \textbf{28}, 139--145 (1989).

\bibitem{Hajian:07}
A.~R. Hajian, B.~B. Behr, A.~T. Cenko, R.~P. Olling, D.~Mozurkewich, J.~T.
  Armstrong, B.~Pohl, S.~Petrossian, K.~H. Knuth, R.~B. Hindsley, M.~Murison,
  M.~Efroimsky, R.~Dantowitz, M.~Kozubal, D.~G. Currie, T.~E. Nordgren,
  C.~Tycner, and R.~S. McMillan, \enquote{Initial results from the {USNO}
  dispersed fourier transform spectrograph,} {\protect\JournalTitle{The
  Astrophysical Journal}} \textbf{661}, 616--633 (2007).

\bibitem{Makiwa:13}
G.~Makiwa, D.~A. Naylor, M.~Ferlet, C.~Salji, B.~Swinyard, E.~Polehampton, and
  M.~H.~D. van~der Wiel, \enquote{Beam profile for the {H}erschel-{SPIRE}
  {F}ourier {T}ransform {S}pectrometer,} {\protect\JournalTitle{Appl. Opt.}}
  \textbf{52}, 3864--3875 (2013).

\bibitem{Swinyard:14}
B.~M. {Swinyard}, E.~T. {Polehampton}, R.~{Hopwood}, I.~{Valtchanov}, N.~{Lu},
  T.~{Fulton}, D.~{Benielli}, P.~{Imhof}, N.~{Marchili}, J.~. {Baluteau}, G.~J.
  {Bendo}, M.~{Ferlet}, M.~J. {Griffin}, T.~L. {Lim}, G.~{Makiwa}, D.~A.
  {Naylor}, G.~S. {Orton}, A.~{Papageorgiou}, C.~P. {Pearson}, B.~{Schulz},
  S.~D. {Sidher}, L.~D. {Spencer}, M.~H.~D. v.~d. {Wiel}, and R.~{Wu},
  \enquote{Calibration of the herschel spire fourier transform spectrometer,}
  {\protect\JournalTitle{Monthly Notices of the Royal Astronomical Society}}
  \textbf{440}, 3658--3674 (2014).

\bibitem{Valtchanov:17}
I.~Valtchanov, R.~Hopwood, G.~Bendo, C.~Benson, L.~Conversi, T.~Fulton, M.~J.
  Griffin, T.~Joubaud, T.~Lim, and N.~Lu, \enquote{Correcting the
  extended-source calibration for the herschel-spire fourier-transform
  spectrometer,} {\protect\JournalTitle{Monthly Notices of the Royal
  Astronomical Society}} \textbf{475}, 321–330 (2017).

\bibitem{Glen:21}
J.~Glenn, C.~M. Bradford, E.~Rosolowsky, R.~Amini, K.~Alatalo, L.~Armus, A.~J.
  Benson, T.-C. Chang, J.~Darling, P.~K. Day, J.~L. Domber, D.~Farrah,
  B.~Hensley, S.~J. Lipscy, B.~D. Moore, S.~Oliver, J.~Perido, D.~C. Redding,
  J.~M. Rodgers, R.~Shirley, H.~A. Smith, J.~B. Steeves, C.~E. Tucker, and
  J.~Zmuidzinas, \enquote{{Galaxy Evolution Probe},}
  {\protect\JournalTitle{Journal of Astronomical Telescopes, Instruments, and
  Systems}} \textbf{7}, 1 -- 36 (2021).

\bibitem{Davidson:04}
D.~B. {Davidson}, \enquote{A review of important recent developments in
  full-wave cem for rf and microwave engineering [computational
  electromagnetics],} in \emph{Proc. ICCEA 2004. 2004 3rd International
  Conference on Computational Electromagnetics and Its Applications, 2004.},
  (2004).

\bibitem{BornWolf:99}
M.~Born and E.~Wolf, \emph{Principles of Optics: Electromagnetic Theory of
  Propagation, Interference and Diffraction of Light} (Cambridge University
  Press, 1999), 7th ed.

\bibitem{Goodman:05}
J.~W. Goodman, \enquote{Introduction to fourier optics,}
  {\protect\JournalTitle{Introduction to Fourier optics, 3rd ed., by JW
  Goodman. Englewood, CO: Roberts \& Co. Publishers, 2005}} \textbf{1} (2005).

\bibitem{Goldsmith:98}
P.~F. {Goldsmith}, \enquote{Quasi-optical techniques,}
  {\protect\JournalTitle{Proceedings of the IEEE}} \textbf{80}, 1729--1747
  (1992).

\bibitem{Sullivan:09}
C.~O'Sullivan, J.~A. Murphy, M.~L. Gradziel, J.~Lavelle, T.~Peacocke,
  N.~Trappe, G.~S. Curran, D.~R. White, and S.~Withington, \enquote{{Optical
  modelling using Gaussian beam modes for the terahertz band},} in
  \emph{Terahertz Technology and Applications II,}  vol. 7215 K.~J. Linden,
  L.~P. Sadwick, and C.~M. O'Sullivan, eds., International Society for Optics
  and Photonics (SPIE, 2009), pp. 174 -- 185.

\bibitem{Withington:21}
S.~Withington, \enquote{{F}unctional {A}nalysis of {P}artially {C}oherent
  {G}rating {S}pectrometers,} Manuscript to be published.

\bibitem{Martin:70}
D.~Martin and E.~Puplett, \enquote{Polarised interferometric spectrometry for
  the millimetre and submillimetre spectrum,} {\protect\JournalTitle{Infrared
  Physics}} \textbf{10}, 105--109 (1970).

\bibitem{Wolf:07}
E.~Wolf, \emph{Introduction to the Theory of Coherence and Polarization of
  Light} (Cambridge University Press, 2007), 1st ed.

\bibitem{Wolf:82}
E.~Wolf, \enquote{New theory of partial coherence in the space--frequency
  domain. part i: spectra and cross spectra of steady-state sources,}
  {\protect\JournalTitle{J. Opt. Soc. Am.}} \textbf{72}, 343--351 (1982).

\bibitem{Withington:01}
S.~Withington and G.~Yassin, \enquote{Power coupled between partially coherent
  vector fields in different states of coherence,} {\protect\JournalTitle{J.
  Opt. Soc. Am. A}} \textbf{18}, 3061--3071 (2001).

\bibitem{Withington:07}
S.~Withington and G.~Saklatvala, \enquote{Characterizing the behaviour of
  partially coherent detectors through spatio-temporal modes,}
  {\protect\JournalTitle{Journal of Optics A: Pure and Applied Optics}}
  \textbf{9}, 626 (2007).

\bibitem{Withington:04}
S.~Withington, M.~P. Hobson, and R.~H. Berry, \enquote{Representing the
  behavior of partially coherent optical systems by using overcomplete basis
  sets,} {\protect\JournalTitle{J. Opt. Soc. Am. A}} \textbf{21}, 207--217
  (2004).

\bibitem{Ozaktas:02}
H.~M. Ozaktas, S.~Y\"{u}ksel, and M.~A. Kutay, \enquote{Linear algebraic theory
  of partial coherence: discrete fields and measures of partial coherence,}
  {\protect\JournalTitle{J. Opt. Soc. Am. A}} \textbf{19}, 1563--1571 (2002).

\bibitem{Murphy:87}
J.~Murphy, \enquote{Distortion of a simple gaussian beam on reflection from
  off-axis ellipsoidal mirrors.} {\protect\JournalTitle{Int J Infrared Milli
  Waves}} \textbf{8}, 1165–1187 (1987).

\bibitem{Schweizer:79}
F.~Schweizer, \enquote{Anamorphic magnification of grating spectrographs - a
  reminder,} {\protect\JournalTitle{Publications of the Astronomical Society of
  the Pacific}} \textbf{91}, 149 (1979).

\bibitem{Kano:62}
Y.~Kano and E.~Wolf, \enquote{Temporal coherence of black body radiation,}
  {\protect\JournalTitle{Proceedings of the Physical Society}} \textbf{80},
  1273--1276 (1962).

\bibitem{Chuss:08}
D.~T. Chuss, E.~J. Wollack, S.~H. Moseley, S.~Withington, and G.~Saklatvala,
  \enquote{Diffraction considerations for planar detectors in the few-mode
  limit,} {\protect\JournalTitle{Publications of the Astronomical Society of
  the Pacific}} \textbf{120}, 430--438 (2008).

\bibitem{Day:03}
P.~Day, H.~LeDuc, B.~Mazin, and et~al., \enquote{A broadband superconducting
  detector suitable for use in large arrays,} {\protect\JournalTitle{Nature}}
  p. 817–821 (2003).

\bibitem{Irwin:05}
K.~Irwin and G.~Hilton, \emph{Transition-Edge Sensors} (Springer Berlin
  Heidelberg, Berlin, Heidelberg, 2005), pp. 63--150.

\bibitem{Mahan:18}
J.~R. Mahan, N.~Q. Vinh, V.~X. Ho, and N.~B. Munir, \enquote{Monte carlo
  ray-trace diffraction based on the huygens-fresnel principle,}
  {\protect\JournalTitle{Appl. Opt.}} \textbf{57}, D56--D62 (2018).

\bibitem{Slepian:61}
D.~{Slepian} and H.~O. {Pollak}, \enquote{Prolate spheroidal wave functions,
  fourier analysis and uncertainty — i,} {\protect\JournalTitle{The Bell
  System Technical Journal}} \textbf{40}, 43--63 (1961).

\bibitem{Landau:61}
H.~J. {Landau} and H.~O. {Pollak}, \enquote{Prolate spheroidal wave functions,
  fourier analysis and uncertainty — ii,} {\protect\JournalTitle{The Bell
  System Technical Journal}} \textbf{40}, 65--84 (1961).

\bibitem{Landau:62}
H.~J. {Landau} and H.~O. {Pollak}, \enquote{Prolate spheroidal wave functions,
  fourier analysis and uncertainty — iii: The dimension of the space of
  essentially time- and band-limited signals,} {\protect\JournalTitle{The Bell
  System Technical Journal}} \textbf{41}, 1295--1336 (1962).

\bibitem{Slepian:64}
D.~{Slepian}, \enquote{Prolate spheroidal wave functions, fourier analysis and
  uncertainty — iv: Extensions to many dimensions; generalized prolate
  spheroidal functions,} {\protect\JournalTitle{The Bell System Technical
  Journal}} \textbf{43}, 3009--3057 (1964).

\bibitem{Slepian:78}
D.~{Slepian}, \enquote{Prolate spheroidal wave functions, fourier analysis, and
  uncertainty — v: the discrete case,} {\protect\JournalTitle{The Bell System
  Technical Journal}} \textbf{57}, 1371--1430 (1978).

\bibitem{Kintner:75}
E.~C. Kintner, \enquote{Edge-ringing and fresnel diffraction,}
  {\protect\JournalTitle{Optica Acta: International Journal of Optics}}
  \textbf{22}, 235--241 (1975).

\end{thebibliography}

\end{document}